# Molecular Cross-linking of MXenes: Tunable Interfaces and Chemiresistive Sensing

*Yudhajit Bhattacharjee* [1, *], *Lukas Mielke* [1], *Mahmoud Al-Hussein* [1,2], *Shivam Singh*[# 3,4], *Karen Schaefer*[# 5,6], *Borja Rodriguez-Barea* [7,8], *Qiong Li* [1], *Anik Kumar Ghosh* [1], *Artur Erbe*[7,8], *Carmen Herrmann* [5,6], *Yana Vaynzof* [3,4], *Andreas Fery* [1,9], *Hendrik Schlicke* [1, *]

Y.B., L.M., M.A.H., Q.L., A.K.G., A.F. and H.S. [1]: Leibniz Institute of Polymer Research Dresden, 01069 Dresden, Germany
M.A.H. [2]: Department of Physics, University of Jordan, 11942 Amman, Jordan
S.S., Y.V. [3]: Leibniz-Institute for Solid State and Materials Research Dresden (IFW Dresden), 01069 Dresden
S.S., Y.V.[4]: Chair of Emerging Electronic Technologies, Dresden University of Technology, 01187 Dresden, Germany
K.S., C.H.[5]: Department of Chemistry, University of Hamburg, 22761 Hamburg, Germany
K.S., C.H.[6]: The Hamburg Centre for Ultrafast Imaging (CUI), Hamburg 22761, Germany
B.R.B., A.E[7]: Institute of Ion Beam Physics and Materials Research, Helmholtz-Zentrum Dresden-Rossendorf (HZDR), 01328 Dresden, Germany
B.R.B., A.E[8]: Institute of semiconductors and microsystems, Dresden University of Technology, 01062 Dresden, Germany
A.F.[9]: Chair of Physical Chemistry of Polymeric Materials, Dresden University of Technology, 01062 Dresden, Germany
*: Corresponding Author. #: Equal Contribution

## Abstract

MXenes, a family of 2D transition metal compounds, have emerged as promising materials due to their unique electronic properties and tunable surface chemistry. However, the translation of these nanoscale properties into macroscopic devices is constrained by suitable cross-linking strategies that enable both processability and controlled inter-flake charge transport. Herein, we demonstrate the tunability of interfaces and the inter-layer spacing between $Ti_3C_2T_x$ MXene flakes through molecular cross-linking with homologous diamines. Oleylamine was first used to stabilize MXenes in chloroform, followed by diamine-mediated cross-linking to tune precisely the interlayer spacing. Grazing incidence X-ray scattering (GIXRD/GIWAXS) confirmed the correlation between ligand chain length and inter-layer spacing, which was further supported by Density Functional Theory (DFT) calculations. Furthermore, we investigated the charge transport properties of thin films consisting of these diamine-crosslinked $Ti_3C_2T_x$ MXenes and observed a strong dependence of the conductivity on the interlayer spacing. The dominating charge transport mechanism is variable range hopping (VRH) in accordance with the structure analysis of the films. Finally, we probed chemiresistive vapor sensing in MXene composites, observing pronounced water sensitivity and selectivity, highlighting their potential for use in humidity sensors. Insights into molecular cross-linking and its impact on charge transport open avenues for next-generation MXene-based electronic devices.

## 1. Introduction

MXenes are an intriguing class of two-dimensional (2D) materials based on transition metal carbides, nitrides, and carbonitrides, which have garnered significant interest from researchers due to their exceptional electronic, mechanical, and chemical properties. [1-3] These materials were first identified in the early 2010s and are derived from their precursor MAX phases, a ternary compound with the general formula $M_{n+1}AX_n$ (where M represents a transition metal, A is a group 13 or 14 element, and X stands for carbon or nitrogen). Synthesizing MXenes involves a selective etching process that removes the A-layer from MAX phases, typically using fluoride-containing acidic solutions like hydrofluoric acid (HF). This results in layered 2D structures composed of transition metal carbide, nitride or carbonitride layers with various surface terminations, including -OH, -F, and -O.[4, 5] The general formula of MXenes, $M_{n+1}X_nT_x$, reflects their ability to be modified, where M indicates a transition metal, X refers to carbon or nitrogen, and $T_x$ denotes the functional groups on their surfaces. This structural versatility allows MXenes to display a wide range of customized properties, making them highly promising candidates for applications in energy storage, electromagnetic shielding, catalysis, water purification, and sensing technologies.[6-9] Among the variety of MXenes, $Ti_3C_2T_x$ has emerged as one of the most studied due to its combination of metallic conductivity, hydrophilicity, and adaptable surface chemistry.[8, 10]

The remarkable electronic properties of $Ti_3C_2T_x$ MXene stem from the delocalized electrons within the transition metal layers, coupled with the adjustable impact of surface functional groups.[11] It exhibits metallic conductivity, positioning it as a promising candidate for use in flexible electronics, supercapacitors, and electromagnetic interference shielding.[6, 12] Additionally, the inherent hydrophilicity resulting from surface terminations improves compatibility with aqueous systems, widening its applicability in environmental and biomedical fields. However, in terms of device integration, problems arise from limited solubility in nonpolar solvents that are valuable for ink-based device fabrication methods.[13]

Research into charge transport mechanisms within $Ti_3C_2T_x$ MXene has garnered considerable interest, as these mechanisms are critical for the material's performance across diverse applications such as energy storage, catalysis, and sensing. Device integration of MXenes typically involves films composed of multi-flake and composite systems. At the intra-flake level, charge transport is primarily driven by band-like conduction, supported by a high density of delocalized electrons and minimal carrier-phonon scattering. The presence of large Fröhlich polarons within $Ti_3C_2T_x$ plays a vital role, as these quasi-particles arise from the interaction between charge carriers and lattice vibrations, resulting in altered effective masses and efficient intra-flake charge transport.[14] The ability to manipulate the surface chemistry of $Ti_3C_2T_x$ MXene introduces additional complexity and potential for optimizing its charge transport characteristics. Functional groups like -OH and -F can modify the material's electronic band structure, consequently affecting carrier mobility and conductivity.[11, 14] This tunability is particularly beneficial for applications requiring precise control over the material's electrical and chemical behavior. However, it can also present challenges, as certain surface terminations may increase carrier scattering or disrupt interlayer coupling, leading to decreased overall conductivity.

Conversely, charge transport becomes more intricate at the inter-flake level, relying on thermally activated hopping processes.[14] In composites consisting of multiple MXene flakes, inter-flake transport often dominates overall charge transport, acting as a critical bottleneck. Therefore, controlling, tuning, and enhancing inter-flake transport is of great interest for optimizing electrical properties in MXene-based systems. This dual-mode transport, i.e., band-like conduction within flakes and hopping between them, highlights the delicate interplay between structural order, interlayer spacing, and the nature of surface functionalization.

Recent research in surface functionalization techniques has started to explore the promise of organic ligands in adjusting the properties of $Ti_3C_2T_x$ MXenes.[15, 16] By bonding organic molecules onto MXene surfaces, researchers can tailor hydrophilicity and chemical stability.[17, 18] The relationship between ligand structures and charge transport pathways in MXene systems remains an unexplored research frontier with important implications for developing future high-performance MXene-based devices. Furthermore, $Ti_3C_2T_x$ MXene showcases remarkable potential in the realm of sensing technologies. Its elevated surface area, electrical conductivity, and tunable surface chemistry position it as a promising candidate for detecting vapors and gases.[19-21] Here, the interaction between target analytes and MXene-based materials can induce measurable alterations in electronic properties, enabling their use as resistive sensors.[22]

In this work, we introduce a modular, molecular approach for stabilization and cross-linking of $Ti_3C_2T_x$ MXene to overcome existing limitations (e.g., poor solubility in nonpolar solvents and limited control over inter-flake interfaces in deposited films). This was achieved by the synthesis of an amine-stabilized MXene dispersion in nonpolar solvents, which was subsequently deposited via the addition of diamines, leading to controlled variation of interlayer spacing and tunable charge transport properties. While a prior study has employed diamine intercalation to manipulate MXene interlayer spacing for battery-related applications[23], these efforts focused on optimizing ion transport in aqueous systems. In contrast, the present work makes use of the $Ti_3C_2T_x$-amine interactions in a modular, surface engineering-based approach that enables processability and subsequently enables fabrication of hybrid organic-inorganic MXene-diamine systems with tunable interfaces and charge transport characteristics.

Specifically, in the first step, we introduce oleylamine (OAm) as a stabilizer suitable for delamination and stabilizing $Ti_3C_2T_x$ MXene flakes. Second, we apply α,ω-diaminoalkane (DA) ligands of varying chain lengths: 1,6-diaminohexane (6DA), 1,8-diaminooctane (8DA), and 1,10-diaminodecane (10DA) for cross-linkage. Applying the homologous molecules, we tuned the inter-layer spacing as indicated by GIWAXS measurements and supported via DFT calculations. The tunable inter-layer spacing further has a strong effect on the electrical conductivity of thin films from MXene-DA composites, reflected in a drastic decrease in conductivity, along with an increase in the length of the cross-linking amine backbones. Additionally, the MXene-DA systems demonstrate remarkable performance as vapor sensors with strong selectivity towards water. Their pronounced selectivity for water is attributed to the interactions between water molecules and the MXene surface, functionalized with amine groups. Herein, the reported approach offers a molecular-level framework for engineering properties such as charge transport in hybrid organic-inorganic MXene films.

## 2. Results and Discussion

## 2.1 Synthesis of a Stable, Functionalized MXene Dispersion

We report the synthesis of a highly stable dispersion of $Ti_3C_2T_x$ MXene in chloroform ($CHCl_3$) through oleylamine (OAm) functionalization. The preparation process is schematically illustrated in Figure 1a. First, commercially supplied pristine $Ti_3C_2T_x$ MXene powder (Figure 1b) was dispersed in chloroform (10 mg/mL), and oleylamine (55 mM) was added. The resulting dispersion was sonicated for one hour and kept stirring for around 48 h at room temperature. After that, the dispersion was again sonicated for about an hour before purification. The MXene-OAm dispersion was subsequently purified in a series of sonication-precipitation steps (section S2) to remove residual larger grains of non-delaminated MXene. To evaluate the colloidal stability of the resulting functionalized $Ti_3C_2T_x$ dispersion, we monitored its macroscopic appearance over time (SI, Figure S2). The dispersion remained stable without noticeable sedimentation for up to 720 hours (30 days), demonstrating the effectiveness of OAm functionalization in preventing re-stacking and aggregation via inter-flake hydrogen bonding.[24] Further, stability was evidenced by the observation that the conductivity of layers (Section 2.6) fabricated from a thirty-day-old dispersion show only negligible deviations from measurements of films initially fabricated from the dispersion right after its preparation (SI, Figure S4). We attribute the stability of the obtained dispersions to the surface functionalization of $Ti_3C_2T_x$ with the long-chain OAm molecules, which expose a hydrophobic chain that enables solubility in $CHCl_3$ (Figure 1a).

The morphology of the OAm functionalized $Ti_3C_2T_x$ flakes was analyzed using electron microscopy and atomic force microscopy (AFM). For comparison, Figure 1b shows a scanning electron microscopy (SEM) image of commercially supplied pristine $Ti_3C_2T_x$ material, revealing the typical layered structure of MXene flakes. Here, the layer thickness of individual, initial $Ti_3C_2T_x$ flakes, prior to OAm functionalization, was 1 nm, based on AFM measurements on flakes delaminated via ultrasonication (SI, Figure S1). The stable dispersion obtained after OAm functionalization was drop-cast onto carbon-coated grid substrates for transmission electron microscopy (TEM) and wafer substrates for atomic force microscopy (AFM). TEM analysis (Figure 1c) highlights the thin, delaminated nature of the flakes with lateral dimensions in the submicron range. AFM measurements provide insight into the thickness of the functionalized MXene flakes, confirming their successful exfoliation into few-layer structures. The height profile indicates that the OAm functionalized MXene individual flakes have a thickness of ~2.7 nm, consistent with successful surface modification and effective dispersion stabilization. Considering the expected thickness for individual layers of $Ti_3C_2T_x$, the double-sided surface functionalization with OAm contributes to the overall thickness observed in AFM.

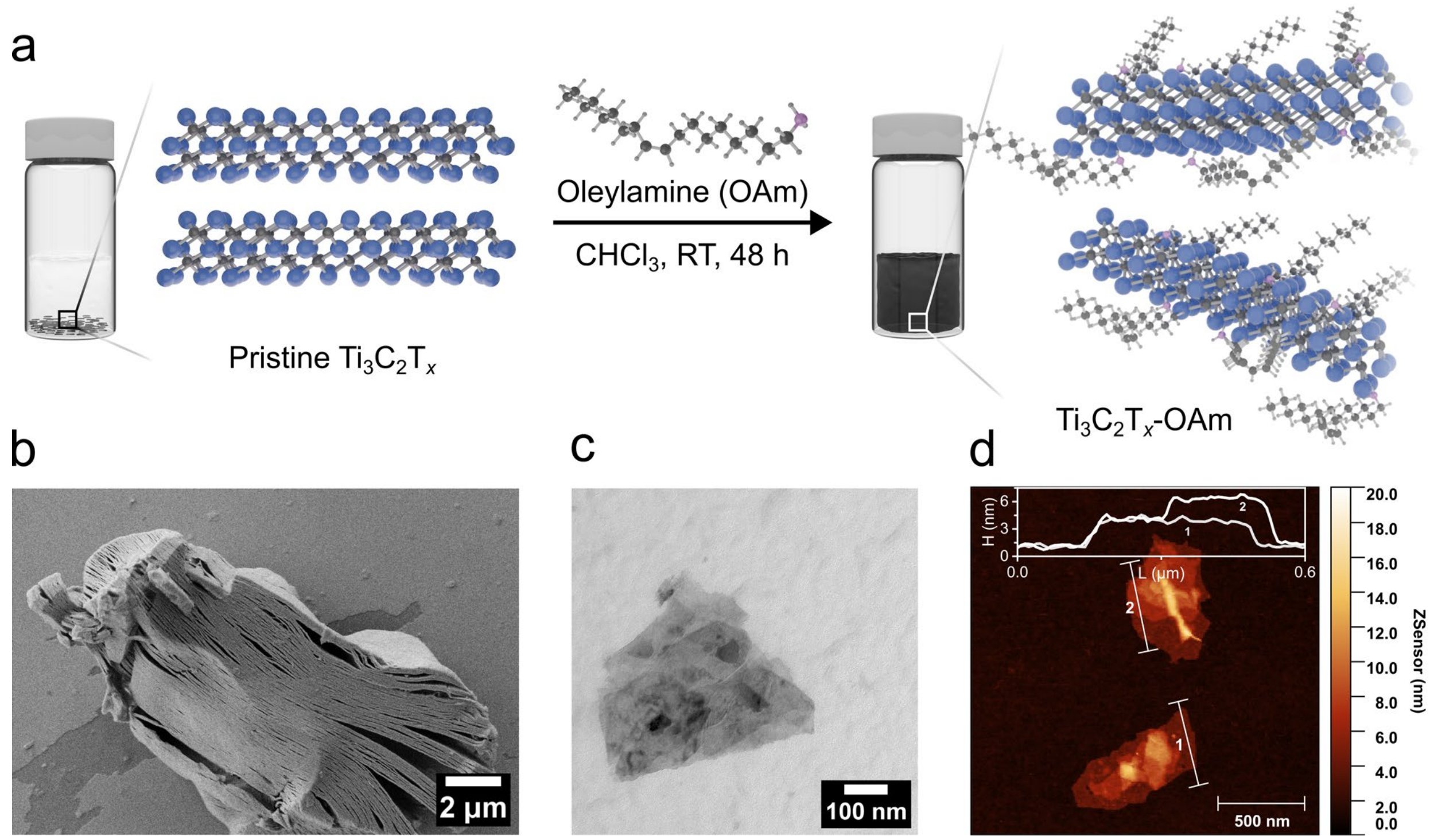

**Figure 1.** (a) Schematic representation of MXene functionalization using OAm, (b) Scanning electron micrograph of pristine $Ti_3C_2T_x$, (c) Transmission electron micrograph of $Ti_3C_2T_x$-OAm flakes, and (d) Atomic force micrograph of $Ti_3C_2T_x$-OAm flakes. (Inset: height profile sampled along $Ti_3C_2T_x$-OAm flakes)

## 2.2 Cross-Linking of MXenes via Ligand Exchange:

Following the synthesis of a stable dispersion of MXene flakes via steric stabilization through functionalization with OAm, we deposited the material onto interdigitated electrode (IDE) structures via spin coating to perform current-voltage measurements. As expected, the film sample made from the $Ti_3C_2T_x$-OAm dispersion exhibited poor current output (Figure S3). We expect these low currents to be due to the long-chain OAm molecules located on the $Ti_3C_2T_x$ MXene surface, which lead to ineffective electronic coupling across the interfaces between flakes. To demonstrate the tunability of the inter-sheet interfaces, we employ cross-linking of MXene flakes using suitable diamines as a bifunctional linker. Therefore, α,ω-diaminoalkanes (6DA, 8DA and 10DA) were introduced into purified, stable dispersions of OAm-stabilized $Ti_3C_2T_x$ flakes. Following the addition of the cross-linkers in ethanol solution, precipitation of black material was clearly visible (Figure 2a). The precipitation indicates a shift from a colloidally stable dispersion to a crosslinked, aggregated state. This behavior can be attributed to the exchange of surface-bound OAm molecules with DA cross-linkers, inducing aggregation and the formation of three-dimensional networks of flakes (Figure 2b). This observation is similar to aggregation mechanisms observed in the cross-linking of gold nanoparticles by adding dithiols to colloidally stable dispersions.[25] Notably, no precipitation occurred when only ethanol was added to the dispersion of OAm-stabilized $Ti_3C_2T_x$ flakes, indicating the crucial role of α,ω-diaminoalkanes as cross-linkers.

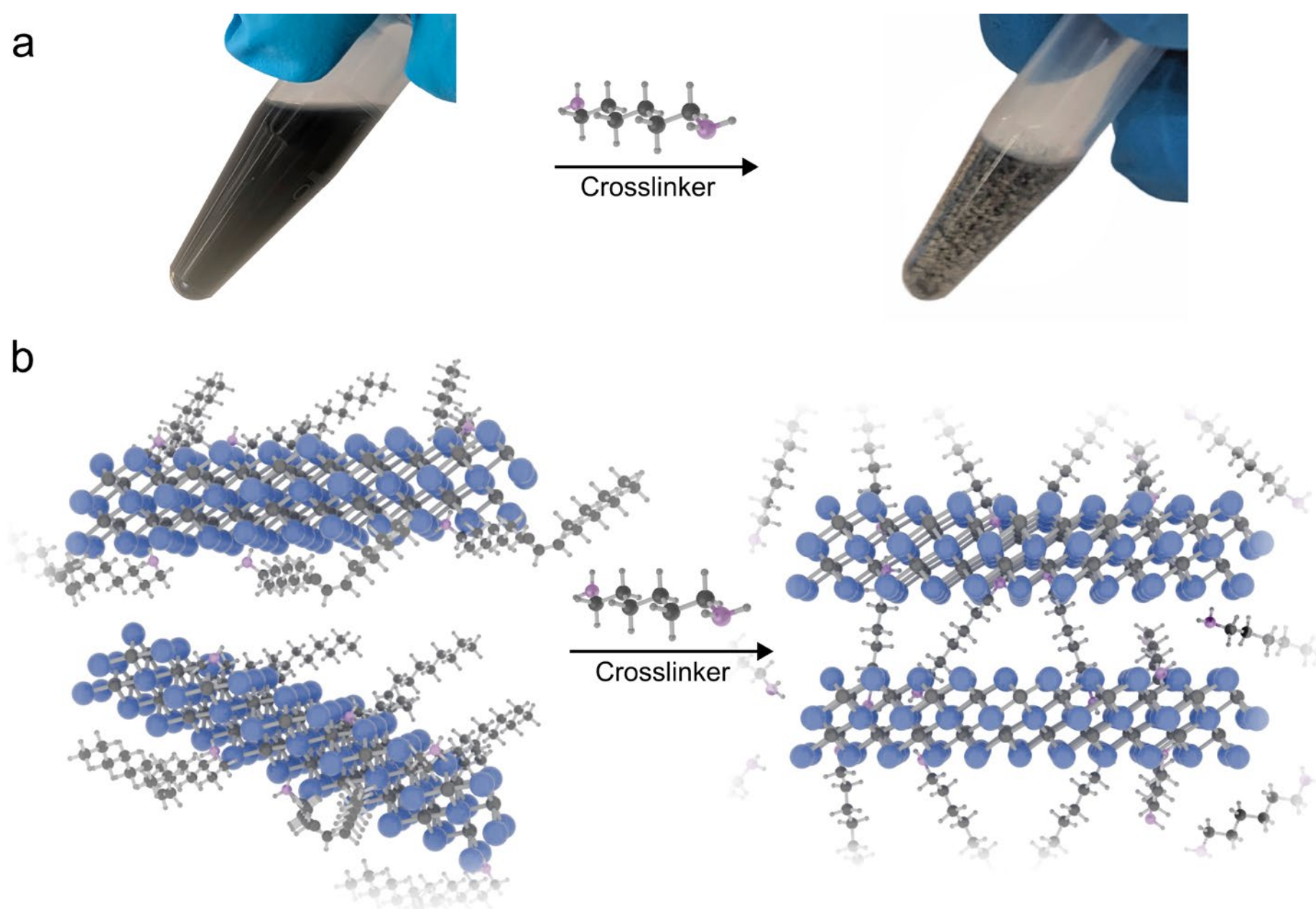

**Figure 2.** (a) Photographs showing a stable dispersion of OAm stabilized MXenes in chloroform before (left) and after (right) the addition of DAs. (b) Sketch depicting the expected nanostructure of the material, i.e., dispersed, OAm-stabilized MXene flakes (left) and stacks of DA cross-linked MXene flakes (right).

## 2.3 Grazing Incidence X-Ray Analysis of Cross-Linked MXenes

To gain further insights into the molecular packing and structural variations between the differently functionalized MXenes, grazing incidence x-ray diffraction and wide-angle x-ray scattering (GIXRD and GIWAXS) measurements were carried out. Figure 3b shows the GIXRD scattering curves of films deposited from dispersions from pristine $Ti_3C_2T_x$, OAm-stabilized $Ti_3C_2T_x$, and DA cross-linked $Ti_3C_2T_x$, drop-cast onto silicon substrates. As can be seen, the pristine $Ti_3C_2T_x$ exhibited a series of Bragg peaks corresponding to the (002), (004), and (008) reflections of $Ti_3C_2T_x$ layers. Interestingly, the scattering data from OAm-stabilized $Ti_3C_2T_x$ exhibited a different (00ℓ) series of reflections, up to the $4^{th}$ order, at a lower scattering vector q*, which is typical for highly ordered layered structures. In contrast, the curves of the DA cross-linked $Ti_3C_2T_x$ films showed a relatively broad reflection at scattering vector q* and a weaker reflection at almost 2q*, indicating variation in the packing of the MXene layers upon functionalization with OAm and different DA molecules.

For more quantitative analysis, the peak positions and the full width at half maximum (FWHM) of the (002) peak of $Ti_3C_2T_x$ and the first reflection at q* for the functionalized films were extracted to determine the corresponding interlayer spacings and coherence lengths, respectively (Table 1). It is noteworthy that functionalization with DA molecules led to a substantial increase in both the interlayer spacing (from 10.0 Å to 45.2 Å) and the coherence length (from 97.55 Å to 224.95 Å) compared to corresponding values of the pristine $Ti_3C_2T_x$ MXene film. These observations indicate that the OAm functionalization improved the stacking of the MXene layers in dried films significantly due to the insertion of the OAm molecules as interlayers. On the other hand, cross-linking with the DA molecules reduced the interlayer spacing considerably compared to OAm functionalized $Ti_3C_2T_x$. It is also observed that using longer DA molecules leads to an increase in the interlayer spacing. This qualitative trend is in agreement with the earlier study on DA intercalation.[23] Also, the coherence length gradually increased with increasing the length of the DA molecules from 30 Å for 6DA to 53 Å for the 10DA, suggesting better packing of the layers.

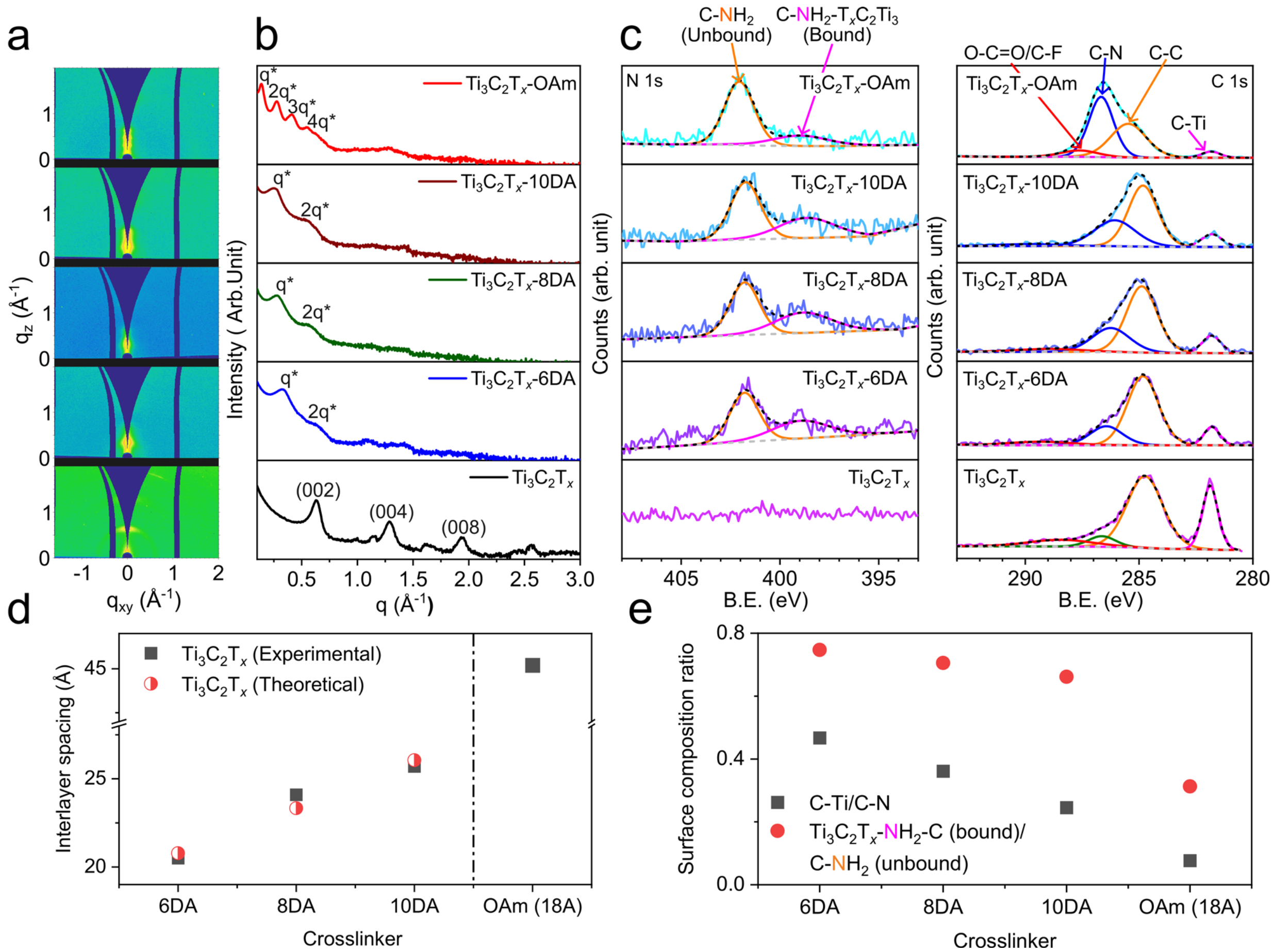

**Figure 3.** (a) GIWAXS patterns of films, drop-cast onto silicon substrates from dispersions of pristine $Ti_3C_2T_x$, $Ti_3C_2T_x$ cross-linked with 6DA, 8DA and 10DA, as well as OAm-stabilized $Ti_3C_2T_x$. (b) Corresponding GIXRD curves. (c) The core-level spectra of N 1s and C 1s for pristine $Ti_3C_2T_x$ and $Ti_3C_2T_x$ cross-linked with 6DA, 8DA, 10DA, and OAm-stabilized $Ti_3C_2T_x$. The black dotted line is the envelope, the grey dotted line depicts the baseline, and the green line in the lower bottom panel of the C 1S spectrum is the C–O signal. (d) Experimentally determined (GIWAXS) and theoretically calculated (DFT) interlayer spacing of $Ti_3C_2T_x$ flakes, cross-linked with 6DA, 8DA and 10DA and experimentally determined interlayer spacing of $Ti_3C_2T_x$ flakes stabilized with OAm. For GIWAXS analysis, the materials were drop-cast from their solutions. (e) Surface composition ratio of functionalized $Ti_3C_2T_x$ with various cross-linkers (6DA, 8DA, 10DA, and OAm-stabilized $Ti_3C_2T_x$). The plot compares the relative ratio of C-Ti/C-N (black squares; from C 1s spectra) and bound to unbound $NH_2$ (red circles; from N 1s spectra) with various cross-linkers.

GIWAXS measurements were also performed to characterize the orientation of the $Ti_3C_2T_x$ layers relative to the substrate. As shown in Figure 3a, the GIWAXS pattern of $Ti_3C_2T_x$ showed an out-of-plane arc-shaped (002) scattering peak, indicating that a large population of the $Ti_3C_2T_x$ layers are oriented in parallel to the substrate. The parallel orientation became even more pronounced for the OAm and DA films, as indicated by the more intensified q* reflections along the out-of-plane direction $q_z$. Moreover, the lateral packing of the DA molecules remained amorphous, as no in-plane Bragg peaks were observed in the GIWAXS patterns of all functionalized films.

**Table 1:** Experimentally determined inter-layer spacing (center-to-center distance) *d* and coherence length $\xi$ obtained from GIXRD/GIWAXS analysis of pristine, 6DA, 8DA and 10DA cross-linked as well as OAm stabilized $Ti_3C_2T_x$. The lower part of the table contains the theoretically calculated (DFT) inter-layer spacing of 6DA, 8DA, and 10DA cross-linked MXene layers. Calculations were conducted taking into account differently terminated MXene structures ($Ti_3C_2(OH)_2$, $Ti_3C_2F_2$ and $Ti_3C_2T_x$).

| **Experiment (GIXRD)** | | | |
|---|---|---|---|
| Species | Ligand/linker | *d* (experiment) (Å) | ξ (Å) |
| $Ti_3C_2T_x$ (neat) | - | 10.0 | 97.55 |
| $Ti_3C_2T_x$ | 6DA | 20.5 | 36.93 |
| | 8DA | 24.1 | 39.62 |
| | 10 DA | 25.7 | 50.10 |
| | 18A | 45.2 | 224.95 |
| **Theory (DFT)** | | | |
| Species | Ligand/linker | *d* (DFT) (Å) | |
| $Ti_3C_2(OH)_2$ | 6DA | 21.04 | _ |
| | 8DA | 23.70 | |
| | 10DA | 26.08 | |
| $Ti_3C_2F_2$ | 6DA | 21.06 | |
| | 8DA | 22.93 | |
| | 10DA | 25.32 | |
| $Ti_3C_2T_x$ ($Ti_3C_2F(OH)$) | 6DA | 20.78 | |
| | 8DA | 23.33 | |
| | 10DA | 26.06 | |

## 2.4 First-Principles Calculations

### 2.4.1 Adsorption of Amine Ligands on MXene Surfaces

To improve our understanding of how amine ligands are adsorbed on $Ti_3C_2T_x$ surfaces, density functional theory (DFT) calculations were performed. During the etching process of the MAX phase in the synthesis of MXenes, surface terminations (T), including -F and -OH, are typically introduced to the surfaces of the MXene layers. We considered three different types of layers in our calculations: (a) a $Ti_3C_2(OH)_2$ layer, (b) a $Ti_3C_2F_2$ layer, and (c) a $Ti_3C_2T_x$ (i.e., $Ti_3C_2F(OH)$) layer, where -F and -OH groups were distributed on the surface [SI, Figure. S5 (a)-(b)]. Details on the preparation of the structures and on the computational method are provided in SI, Section S6.1 in the Supporting Information and in the computational methods section.

For each of the three surface terminations, an amine ligand was adsorbed on the surface. The adsorption was investigated using a butyl amine ligand instead of the longer-chained 6DA to 10DA molecules investigated experimentally to reduce computational costs.

Adsorption energies for all resulting optimized structures are summarized in the SI, Table S1. The interactions between the amine ligand and the terminal groups (-OH and -F) in the $Ti_3C_2(OH)_2$, the $Ti_3C_2F_2$, and the $Ti_3C_2T_x$ (i.e., $Ti_3C_2F(OH)$) structures show relatively weak interactions (adsorption energies of -33.8, -21.2, and -51.4 kJ/mol) corresponding to hydrogen bonding. In addition to the bonding energies, these optimizations gave us information on the preferred binding atomic arrangements, which were then used as input in the calculations on the MXene monolayers discussed below.

### 2.4.2 Center-to-Center Distances of MXene Monolayers

Another question that cannot be answered easily through experimental techniques is the arrangement of DA ligands between the MXene monolayers. Specifically, it remains unclear whether these ligands act as a bridge between the monolayers or if they are solely connected to one of the surfaces. To gain first insights into this question, two MXene layers, separated by a single α,ω-diaminoalkanes ligand (6DA, 8DA, or 10DA), were investigated theoretically for each layer type ($Ti_3C_2(OH)_2$, $Ti_3C_2F_2$, and $Ti_3C_2F(OH)$). Details on the preparation of the structures can be found in the Supporting Information (SI, Section S6.2). We applied periodic boundary conditions to effectively model MXene monolayers (Figure 4a) while keeping computational times short. It should be noted that the periodic replication of the cell introduces an artificial ordering of the -F and -OH groups, which is unlikely to occur in the actual material. However, at this stage, we assume that this simplification does not significantly influence the fundamental properties of the monolayer. In the periodic system, the distance between the amine ligand within the unit cell and its nearest neighboring ligand is approximately 6.2 Å. This serves as our initial estimate for the ligand density in the MXene monolayer, which remains experimentally unknown.

Experimentally, the distances from the center of one monolayer to the center of another are evaluated. These measurements are more easily accessible than e.g., surface-to-surface distances, owing to the surface non-uniformity of the MXene layers. To compare experimental and theoretical distances directly, we extract the theoretical distances from the center of the first to the center of the second monolayer (see Figure 4b for a representation of how the center-to-center distance was evaluated). Details on the optimization protocol can be found in the computational methods section. The center-to-center distances of the structures with the lowest energy for each of the three investigated ligands are summarized in Table 1. For a given ligand length, e.g., 6DA, the center-to-center distances between the two layers are very similar for the $Ti_3C_2(OH)_2$, $Ti_3C_2F_2$ and $Ti_3C_2F(OH)$ cases (21.04, 21.06, and 20.78 Å). The calculated center-to-center distances can be compared directly to those obtained experimentally (cf. Table 1 and Figure 4b,c). A distance of 20.5 Å was obtained from GIWAX for the 6DA ligand, which is closely aligned with the distance obtained for $Ti_3C_2F_1(OH)_1$ (20.78Å) and only about 1 Å lower than the distances obtained for the $Ti_3C_2(OH)_2$ (21.04 Å) and $Ti_3C_2F_2$ (21.06 Å) layers, separated by 6DA in our DFT calculations.

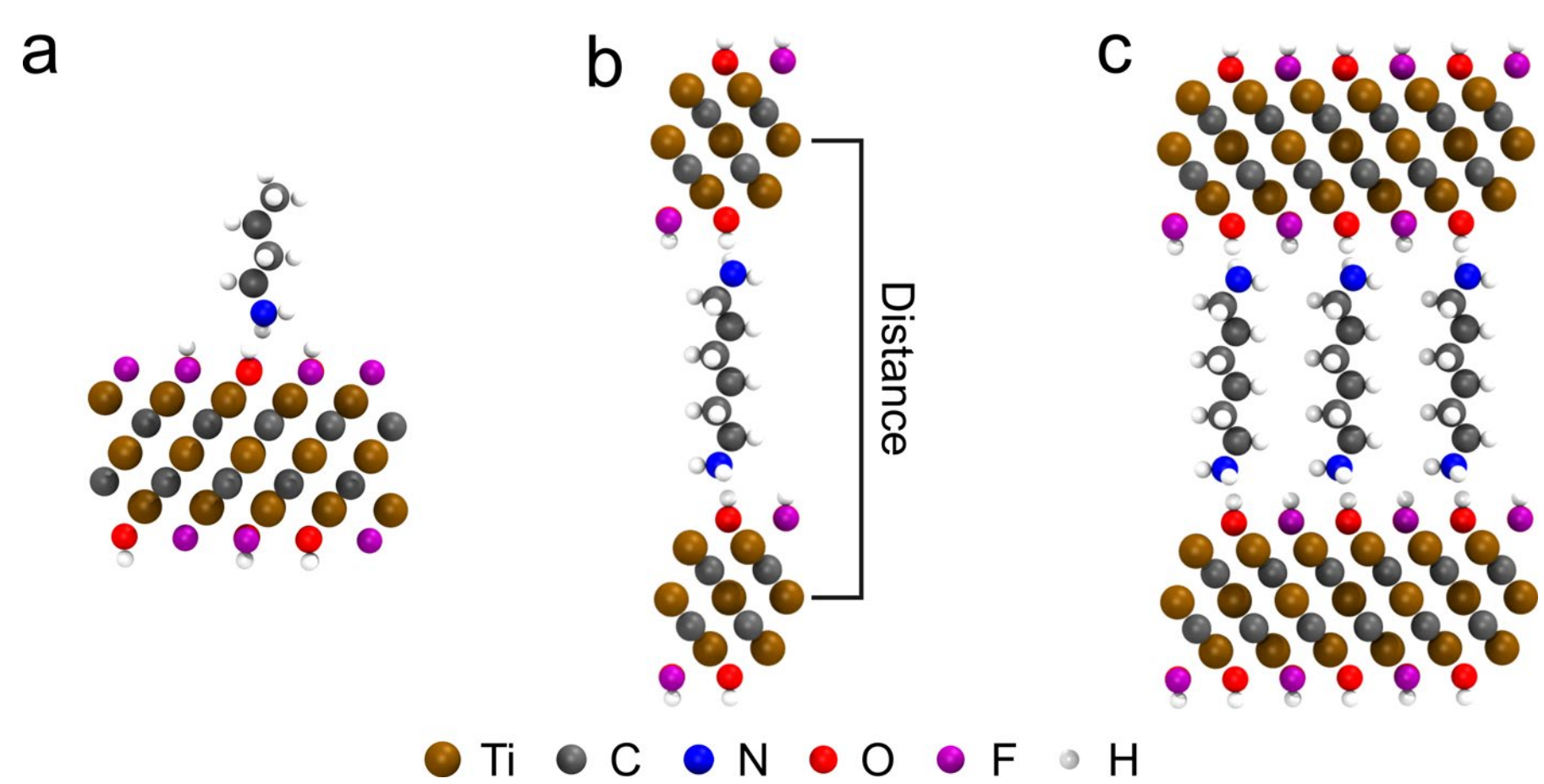

**Figure 4**: Optimized structures of butyl amine ligands adsorbed on (a) $Ti_3C_2F(OH)$ surfaces. (b) Optimized structure of two $Ti_3C_2T_x$ (i.e. $Ti_3C_2F(OH)$) layers that are separated by a 6DA molecule. The distance between the two layers is calculated from the center of one layer to the center of the other layer. Periodic boundary conditions were applied for all four structures (c) Periodically repeated monolayer of the optimized $Ti_3C_2T_x$ unit cell separated by 6DA molecules.

Even though the structures investigated theoretically are simplified, they can give us insights into the structure of these systems. For example, by comparing the calculated distances with experimentally obtained counterparts, we can conclude that the measured center-to-center distance between the layers is small enough such that the DA ligands can potentially form a bridge between the layers. However, it is important to note that while the current non-tilted ligand conformation and ligand density align well with experimental observations, similar center-to-center distances could also be achieved through alternative configurations. These include tilted ligands that do not bridge, a higher or lower ligand coverage, or local fluctuations in ligand density. Future studies can address these issues by incorporating multiple ligands into the unit cell. Additionally, the structures examined here can serve as a foundation for investigating their charge transport properties.

## 2.5 X-Ray Photoemission Spectroscopy

X-ray photoemission (XPS) measurements (Figure 3c) were performed to investigate the surface chemistry and interactions within the modified $Ti_3C_2T_x$ MXene structures. The N 1s spectra show peaks corresponding to N-C species, as is expected for all samples incorporating OAm or the different diamines. We associate the high binding energy peak with unbound DAs that have not interacted with the MXene, while the low binding energy peak at ~399 eV is associated with those that have interacted with the $Ti_3C_2T_x$ MXenes. Our results indicate that this interaction is likely occurring via hydrogen bonding rather than direct Ti-N bonds, as those are expected to result in much lower binding energies of 396.5 eV.[26] Nevertheless, the results indicate the successful interaction between the amine-based cross-linkers and the $Ti_3C_2T_x$ MXene surfaces. Note that the ratio between the peak areas of the amines that interacted with $Ti_3C_2T_x$ MXene and those that did not varies with the length of the cross-linking molecules (Figure 3e). Specifically, 6DA leads to the highest ratio, suggesting that shorter molecules can more effectively interact with the MXene surface than longer ones. A higher binding efficiency was also observed for composites. In the C 1s spectra, contributions from C-Ti, C-N, C-C, and O-C=O/C-F species are evident. The species associated with C-Ti progressively decreases as the length of the amines increases, suggesting that the MXene is surrounded by the amines, thereby attenuating the signal that originates from the MXene flakes themselves. The gradual increase in the C-N component with longer cross-linkers (8DA, 10DA) suggests a higher number of amines being incorporated into the structures, potentially underpinning the above results that indicate higher numbers of free $NH_2$ groups for longer DA or OAm. Interestingly, the Ti 2p and O 1s XPS spectra (SI, Figure S6) indicate a reduction in the appearance of oxidized species (Ti-O) when cross-linkers are used.[27] This serves as additional evidence for the successful interaction between the amine-based cross-linkers to the $Ti_3C_2T_x$ MXene surfaces, thus reducing surface oxidation as is observed in neat MXene. Beyond the stabilization of $Ti_3C_2T_x$ MXenes, such a decrease in the oxidized species associated with MXene might also lead to increased inherent conductivity of the system.

## 2.6 Device Fabrication and Characterization

Thin films from DA cross-linked $Ti_3C_2T_x$ were fabricated on oxidized silicon wafers, equipped with interdigitated electrodes with a channel length (distance between the electrode fingers) of 20 µm and an overlap width of 49 mm. For charge transport characterization, devices were made by a layer-by-layer spin coating procedure (Figure 5a). Within this process, we alternately deposited DA linkers and the OAm stabilized MXene dispersion in a repetitive manner. For obtaining continuous thin films, a total of $n$=16 deposition cycles were conducted. SEM and AFM analysis were employed to analyze film morphology. The SEM micrograph of a 6DA cross-linked $Ti_3C_2T_x$ film depicted in Figure 5b clearly shows the complete coverage throughout the device, and from a topographic AFM scan conducted at a trench within the film, obtained by deliberately scratching it with a cannula, an average film thickness of approximately 24-28 nm was obtained for different $Ti_3C_2T_x$ cross-linked films. (Figure 5c and SI, Section S7)

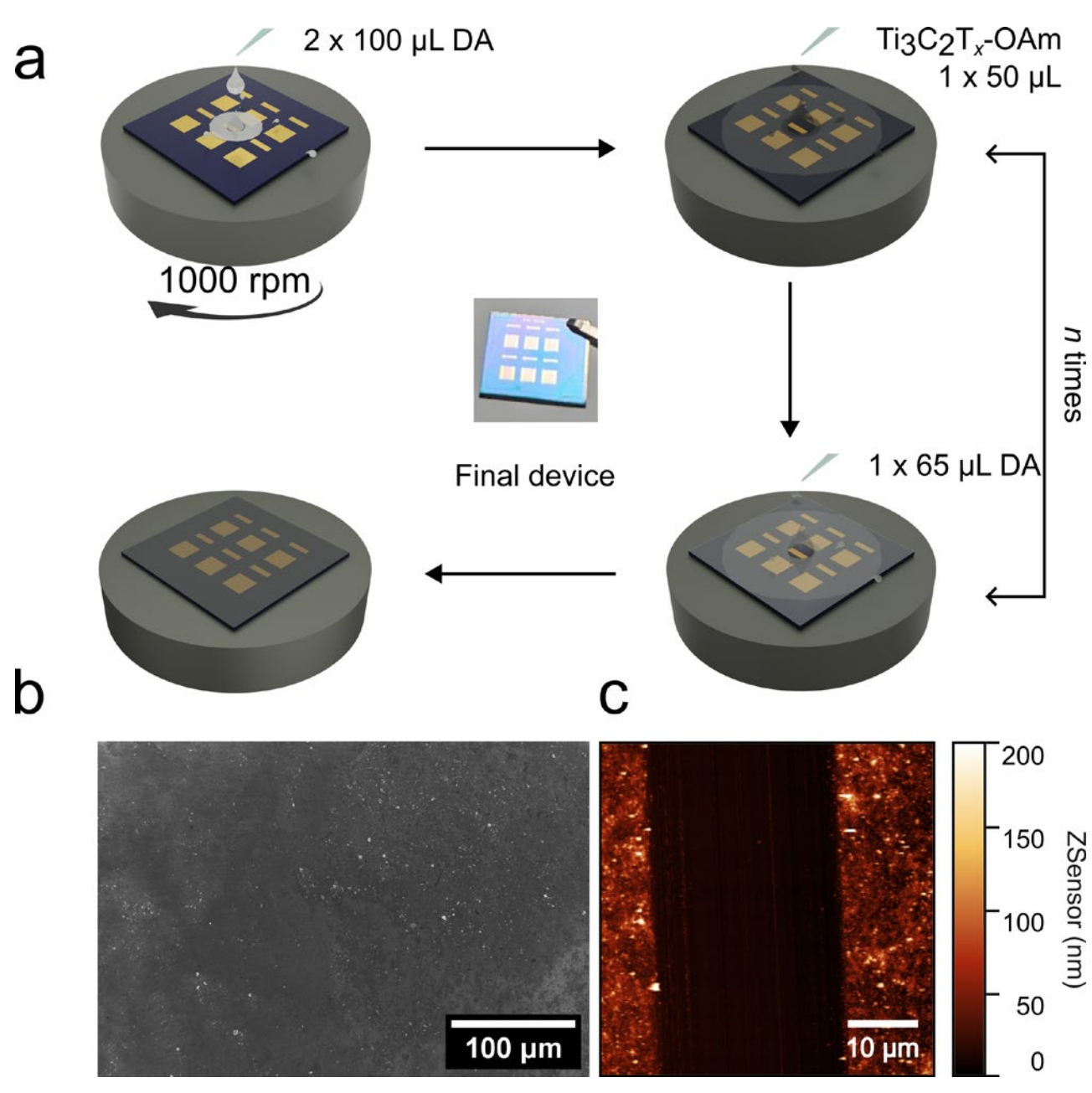

**Figure 5.** (a) Schematic representation of the spin coating-based process for the deposition of DA cross-linked MXene layers. (b) SEM image of a $Ti_3C_2T_x$-6DA film deposited onto a silicon wafer. (c) AFM topography scan on a trench caused by deliberately scratching a $Ti_3C_2T_x$-6DA film for thickness determination.

The I-V characteristics of thin films from $Ti_3C_2T_x$ functionalized with cross-linkers of varying chain lengths (6DA, 8DA, and 10DA) deposited on interdigitated electrodes highlight the achieved control over inter-flake interfaces and the impact of ligand length on charge transport. As seen in Figure 6a (at room temperature and atmospheric pressure) the current decreases drastically with increasing linker length, with 6DA showing the highest current and 10DA the lowest. Overall, for the elongation of the amines' backbone by four methylene units, the conductivity dropped by approximately two orders of magnitude. This trend shows the expected role of DA molecules, determining the overall conductivity of the system. As indicated by scattering experiments described above, shorter ligands, such as 6DA, maintain smaller interlayer distances, which facilitate efficient charge transport via reducing the width of the tunnelling barrier between adjacent MXene layers. Conversely, the longer ligands, such as 10DA, increase the interlayer spacing, leading to a significant reduction in conductivity due to the higher tunneling resistance. This behavior aligns with the expected mechanism of thermally activated hopping for inter-flake charge transport,[14] where larger interlayer separations further hinder the mobility of the charge carriers.

To probe the nature of the charge transport mechanism and its dependence on ligand length, we conducted two sets of temperature-dependent electrical conductivity measurements under high vacuum across the 25–300 K range on different IDE regions of the same 6DA-, 8DA-, and 10DA-functionalized $Ti_3C_2T_x$ films (Figure 6 b,d and SI, Figure S8 and S9). To avoid Joule heating effects, the applied voltage was limited to ±15 mV. The results were analyzed using the Mott variable range hopping (VRH) model, $G(T) \propto \exp\left[-\left(\frac{T_0}{T}\right)^{\frac{1}{(d+1)}}\right]$ assuming three-dimensional transport. For the 6DA-functionalized film, excellent agreement with VRH behaviour was observed across the full temperature range, yielding characteristic $T_0$ values of 5 K and 7 K for the two IDE regions showing lower values than those previously reported for molybdenum-based MXenes.[28] Similarly, 8DA film showed $T_0$ values of 15 K and 19 K, indicating stronger carrier localization with longer ligands in line with the expectation of reduced electronic coupling.[29] 10DA-functionalized film can be fitted to VRH fitting below ~200 K with $T_0$ values of 51 K and 62 K. In contrast to 6DA and 8DA, it displayed a weaker temperature dependence above ~200 K, consistent with a transition to transport governed by distance-limited tunneling (Figure 6c). This suggests that ligand length not only controls interlayer separation but also modulates the dominant transport regime, where for shorter ligands the contribution by VRH is prevalent and for longer ligands tunnelling transport dominates.[30, 31] This insight underscores molecular design as a versatile platform for tuning the electronic properties of $Ti_3C_2T_x$ MXenes by modulating the interlayer spacing. Furthermore, the introduction of more complex molecules, such as conjugated diamines or diamines carrying photoactive groups, could lead to functional composites for advanced optoelectronic applications.

The stability of a device functionalized with 6DA under ambient conditions was probed via monitoring its current-voltage characteristics over several weeks (SI, Figure S10). The experiments show some minor fluctuation of the devices' conductance over the first 17 days, and although a noticeable drop in current was observed, the device still maintained appreciable conductivity after 180 days. The observed stability can be attributed to the chemical bonding between the diamino ligands and the $Ti_3C_2T_x$ surface, effectively preventing the functionalized layers' degradation. Additionally, the inherent robustness of $Ti_3C_2T_x$, both structurally and chemically, plays a crucial role in maintaining the integrity and performance of the devices over extended periods. These results further emphasize the suitability of functionalized MXenes for long-term applications in environmental sensing and electronic devices.

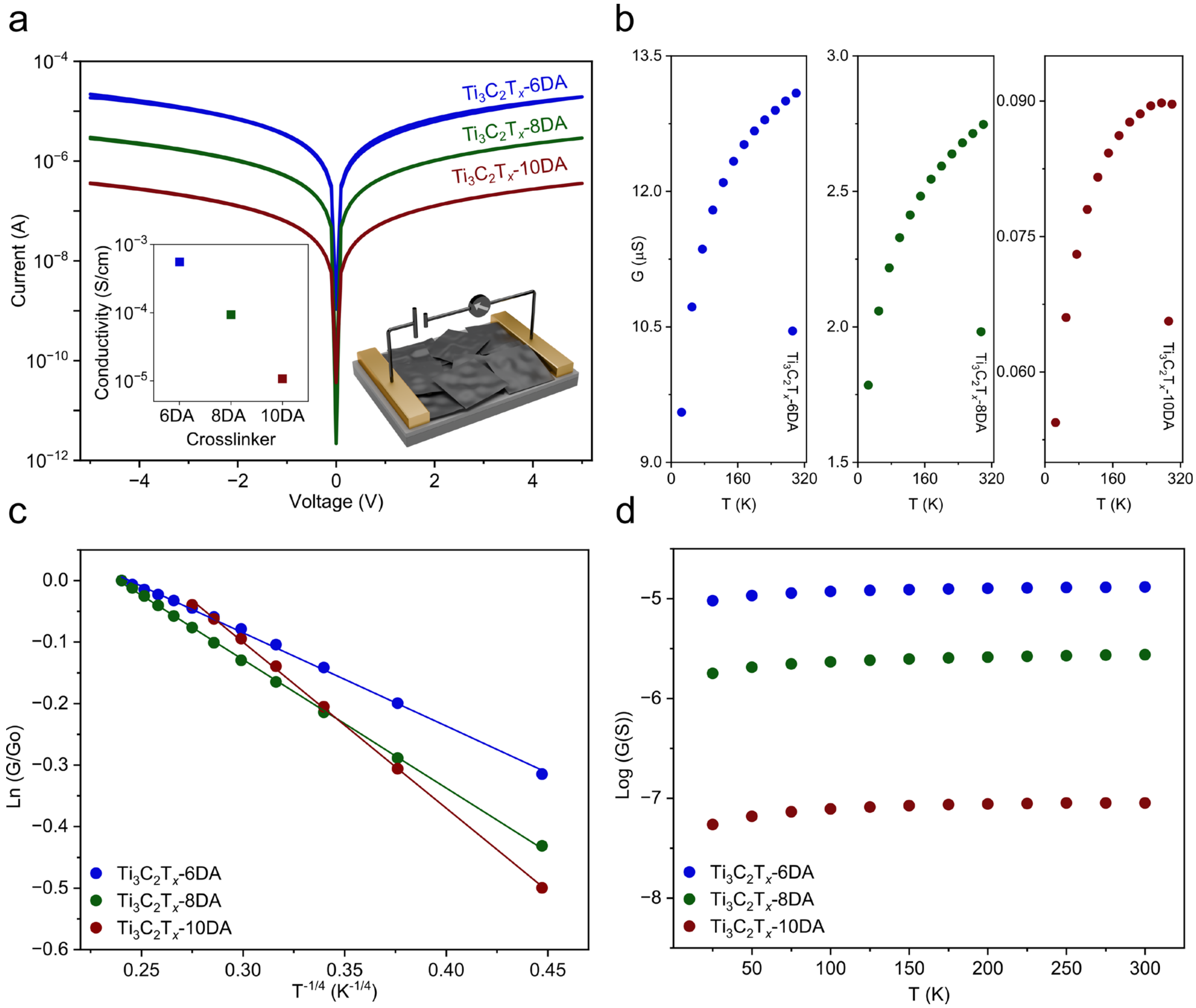

**Figure 6.** (a) Current–voltage (I–V) characteristics of layers from $Ti_3C_2T_x$ cross-linked using different α,ω-diaminoalkanes. (Inset: conductivity vs cross-linker length). (b) shows the conductance as a function of temperature for each ligand system. (c) Normalized natural logarithm of the conductance versus $1/T^{1/4}$ for each functionalized $Ti_3C_2T_x$ film. Solid lines represent a linear fit of the experimental data. Note, the fit for the 10DA-functionalized film is restricted to the 25-200 K range. (d) Logarithmic representation of the conductance for the various functionalized $Ti_3C_2T_x$ films.

## 2.7 Resistive VOC Sensing

Recent studies report that MXenes exhibit pronounced chemiresistive properties, which render them promising candidates for future gas sensors.[22] Due to their hybrid organic/inorganic nature, we expect the hybrid MXene-DA systems investigated in this study to act as sorbents for VOCs and their charge transport characteristics to be sensitively influenced by analyte uptake.

For the investigation of chemiresistive characteristics, substrates equipped with microfabricated electrode structures, coated with neat $Ti_3C_2T_x$ for control and $Ti_3C_2T_x$-DA composite films, as described in the previous section, were placed in a test cell (Figure 7a) and transiently exposed to a set of VOCs in nitrogen with varying concentrations. Meanwhile, the resistance of the device was continuously monitored. Figures 7b and 7c show exemplary resistive responses of a 6DA cross-linked MXene thin film upon exposure to toluene and water vapor with concentrations ranging between 50 and 4000 ppm. Upon exposure, the sensors showed a fully reversible increase in resistance of up to 22.1 % (water vapor, 4000 ppm) with a short response time. Time traces showing the relative resistance changes upon exposure to other VOCs (acetone, ethanol, 1-propanol, methylisobutylketone(MIBK)) at varying concentrations are provided in the Supporting Information (Figure S12). Chemiresistive responses of a $Ti_3C_2T_x$ thin film cross-linked with the longer chain 10DA are depicted in the Supporting Information (SI, Figures S13 and S14). The data indicate an overall similar behavior when compared to the $Ti_3C_2T_x$-6DA film, but a higher noise level due to its higher impedance. For comparison the neat MXene powder was dispersed in ethanol and deposited by drop casting on the electrodes. Conductance measurements showed a strong conductance decrease of these devices within hours. Nevertheless, the resistive sensing ability for water was probed and we did not observe any significant changes in the response (Figure S11), underscoring the importance of the molecular cross-linking strategy.

Figure 7d depicts response isotherms of the $Ti_3C_2T_x$-6DA sensor exposed to acetone, toluene, ethanol, 1-propanol, MIBK and water vapor. The sensor's resistive sensitivity to the different VOCs was extracted from slope fits to the isotherms' low-concentration range (50-400 ppm). Data points represent the response amplitudes after 235 s exposure to the test gas, 5s before stopping the exposure. Further, based on the isotherms' shape, we tentatively fitted a Langmuir-Henry model to the data.[32] The extracted fit parameters are provided in the Supporting Information (Section S12: Table S2 and S3). Overall, the observed sensitivity of $Ti_3C_2T_x$-6DA thin film to different analytes (Figure 7e) is in a similar range as values reported recently for sorption-based chemiresistors from gold nanoparticles cross-linked with varying dithiol molecules.[33]

Interestingly, however, the sensor shows a pronounced sensitivity to water of $1.2 \times 10^{-4}$ $ppm^{-1}$, which exceeds the sensitivities observed for the other probed VOCs by approximately one order of magnitude. We attribute this observation to interactions of water with abundant polar functional groups (-OH and -F) on the MXene surface and with interfaces between them and bound amine groups. Here, strong sorption effects may significantly affect the inter-layer charge transport, resulting in pronounced sensor responses. Bittinger et al. reported a conductive sensitivity of 2.2 %/%RH for graphene oxide/reduced graphene oxide/titania nanocrystals (GO/rGO/TNC)[34], which is comparable to relative resistance changes in this work. A similar resistive sensitivity of ~2%/%RH was observed by Tachibana et al. for cellulose nanofiber/carbon black composites.[35] An overview of characteristics of recently reported humidity sensors based on nanomaterials is provided in the literature.[36]

Our results show that DA cross-linked MXene are a suitable material for VOC sensing. Especially the pronounced sensitivity to water, along with a smoothly increasing response isotherm could render these materials suitable for humidity sensors. Current studies in our group address the investigation of the sensing mechanism to gather insights in the relation between the amount of sorbed analytes and their influence on nanostructure and charge transport of composite materials. Further, we expect a wide tunability of selectivity and sensitivity via altering the composition of the hybrid MXene/diamine systems. This could render these nanocomposites a versatile material platform to form sensor arrays for machine learning-based analyte recognition and quantification.

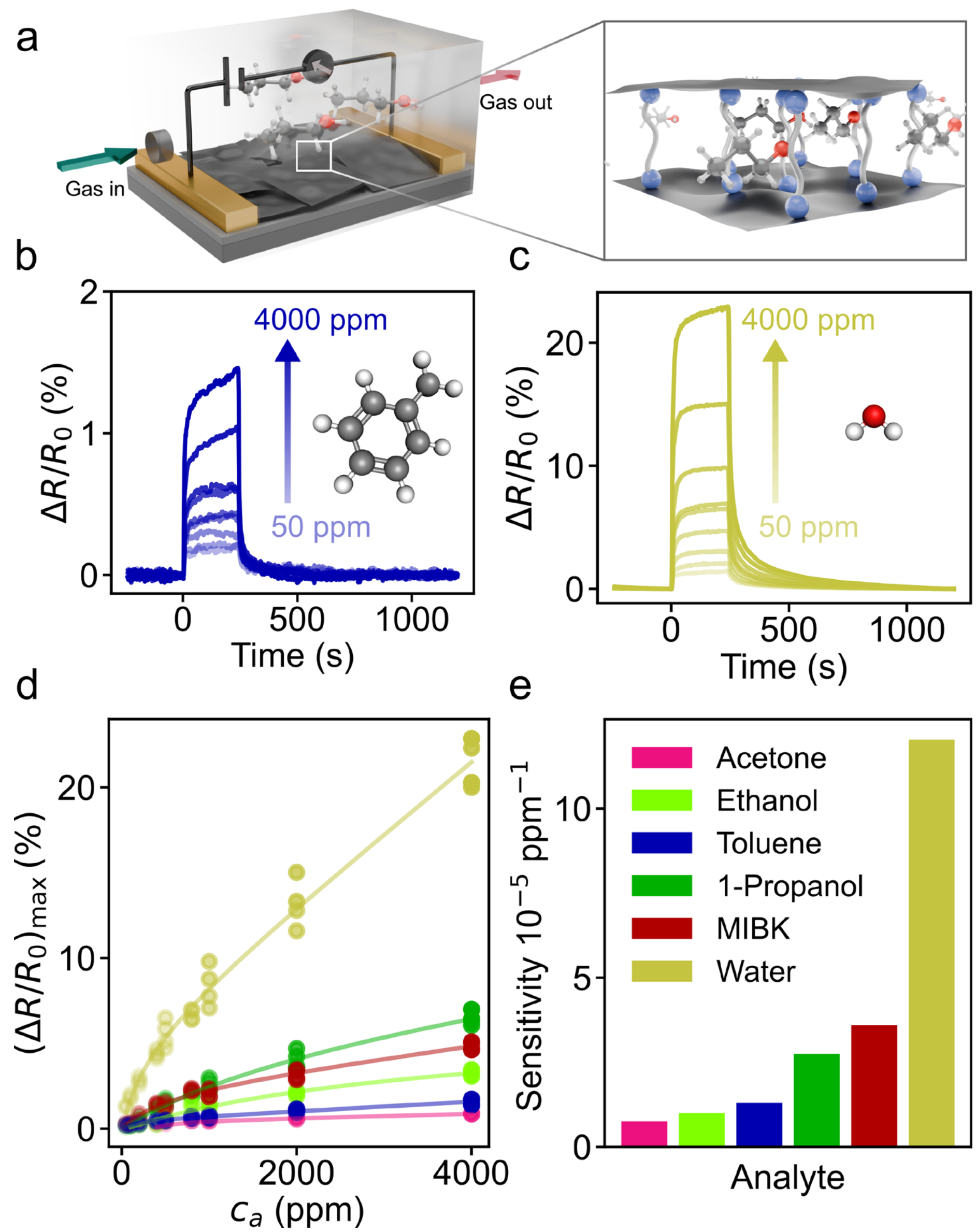

**Figure 7.** (a) Schematic depiction of a chemiresistive gas sensor based on $Ti_3C_2T_x$-6DA composite in a test cell. (b, c) Transient responses of $Ti_3C_2T_x$-6DA chemiresistor to (b) toluene and (c) water at concentrations of 50, 100, 200, 400, 500, 800, 1000, 2000 and 4000 ppm in nitrogen. (d) Response isotherms of the $Ti_3C_2T_x$-6DA-based chemiresistor, exposed to a set of VOCs. The colors correspond to the probed analyte molecules as shown in e. Through lines depict Langmuir-Henry isotherms fitted to the data. (e) Chemiresistive sensitivities of the $Ti_3C_2T_x$-6DA sensor to test gases, extracted from slope fits to data points in a concentration range of 50-400 ppm. For acetone, just the data point of 400 ppm and for ethanol 400, 200 and 100 ppm were considered to be reliable.

## 3. Summary and Conclusion

In summary, this work provides new insights into ligand functionalization and molecular cross-linkage of $Ti_3C_2T_x$ MXenes with amines and diamines, as well as consequences for charge transport in hybrid MXene composite materials. Notably, via surface functionalization with oleylamine, we report the facile one-step synthesis of a MXene dispersion in nonpolar solvent (chloroform), which remains stable for several days. In the following, we introduce homologous α,ω-diaminoalkanes for cross-linking $Ti_3C_2T_x$ layers. Herein, we systematically tuned the interfaces between MXene layers using differently sized cross-linkers. The tunability of the inter-layer spacing was confirmed by X-ray scattering experiments and is in good agreement with DFT calculations. To the best of our knowledge, this cross-linking approach is entirely novel for MXene and 2D materials systems, marking a significant step in the field of MXene and 2D materials research. Layers from diamine cross-linked $Ti_3C_2T_x$ flakes were further deposited onto electrode substrates following a facile spin-coating-based approach. The conductivity of the composite layers was investigated, and a strong dependence on adjustable interlayer distance was observed, based on VRH transport mechanism. This demonstrates the tunability of the electronic coupling between different entities in the functionalized films. Finally, we probed the applicability of the hybrid materials for chemiresistive VOC and water (humidity) sensing. The devices exhibit pronounced, water-selective sensitivity and a smooth response isotherm across a broad concentration range, positioning these hybrid materials as strong candidates for next-generation humidity sensors. The findings presented here not only advance the fundamental understanding of MXene-ligand interactions but also establish a foundation for the development of next-generation MXene-based devices with tailored properties.

## 4. Experimental Section

### Materials

$Ti_3C_2T_x$ MXene powder was obtained from Nanochemzone, Inc., Canada. 1,6-diaminohexane, 1,8-diaminooctane, and 1,10-diaminodecane were purchased from Sigma Aldrich. Oleylamine and ethanol were procured from Acros organics/Thermofischer Chemicals. Thermally oxidized $Si/SiO_2$ (500 nm) wafers were procured from Silicon Materials e.K, Germany. For metal evaporation Ti and Au (99.99% pure) were purchased from Kurt.J. Lesker Company, USA. Acetone was purchased from Fisher Scientific GmbH (≥ 99.8%). 1-propanol was purchased from Grüssing GmbH (IPA, ≥ 99.5%). Toluene was purchased from Thermo Fisher Scientific Inc. (99.85%). Ethanol was purchased from VWR International (≥ 99.8). 4-Methylpentan-2-one was purchased from Sigma-Aldrich (Merck KGaA) (MIBK, ≥ 98.5%). A Q-Pod purification system from Merck KGaA provided ultrapure water (18.2 MΩ·cm). Nitrogen 5.0, used as zero and carrier gas, was purchased from Rießner-Gase GmbH (99.999%).

### Characterization

**SEM & TEM**: TEM images were obtained using a Libra 120 (Zeiss, Germany). Measurements were performed at an acceleration voltage of 120 kV. Samples of dispersed pristine MXenes were diluted to approx. 1 mg/mL with ethanol, and the solution was drop casted and dried overnight on CF200- Cu-50 TEM grids (carbon support film on 200 mesh copper, Electron Microscopy Sciences, USA). Images were analyzed using ImageJ software (1.54n). SEM images were captured at an SE2 detector in a NEON 40 FIB-SEM workstation (Carl Zeiss Microscopy GmbH, Germany). The micrographs were captured at electron high tension (ETH) of 5 kV.

**GIXRD/ GIWAX:** The GIXRD patterns were recorded using XRD 3300 T-T diffractometer (Seifert, GE Sensing & Inspection Technologies, Ahrensburg, Germany) with monochromatic Cu Kα radiation (λ = 1.54 Å). The films were mounted horizontally at the center of a two-circle goniometer and investigated at a fixed incidence angle of 0.4°. The interlayer spacing and coherence length were then calculated from the scattering peak position, $q$, and full width at half maximum, $\Delta q$,according to the following equations: $d=\frac{2\pi}{q}$ and $\frac{2\pi}{\Delta q}$, respectively. The GIWAXS experiments were performed using GANESHA 300XL+ system (SAXSLAB ApS, Lyngby/Denmark). The instrument is equipped with a Pilatus 300K detector, Cu X-ray source operated at 50 kV/0.6 mA (λ = 1.5408 Å), and a three-slit collimation system. An incident angle of 0.2° and a sample-to-detector distance of 102.2 mm were used.

**AFM**: The thickness of the MXene individual flake illustrated in Figure 1d and Figure S1 was studied using a Dimension ICON (Bruker-Nano, USA) in the peak force tapping mode. Silicon nitride sensors SCANASYST-AIR (Bruker, USA) with a nominal spring constant of 0.4 N/m and a tip radius of 2 nm were utilized. In order to achieve high-resolution imaging in air, the peak-force set point was maintained at a value less than 40 mV. The thickness of the cross-linked MXene films illustrated in Figure 5c and Figure S6 b-d was studied using the Dimension ICON (Bruker-Nano, USA) in tapping mode. OTESPA cantilevers (OPUS by MikroMasch) with a nominal spring constant of 26 N/m and a tip radius < 7 nm were utilized. The obtained data were processed using the Gwyddion software and a custom-made Python script.

**XPS**: The samples were transferred to an ultrahigh vacuum chamber (ESCALAB 250Xi, Thermo Scientific) with a base pressure of $1\times10^{-10}$ mbar for XPS measurements. XPS measurement was carried out using an XR6 monochromated Al kα source (hυ = 1486.6 eV) using a spot size of 650 μm. Pass energy of 100 and 20 eV was used for the survey and core level spectra, respectively.

**Room-Temperature Electrical Measurements**: Room-temperature conductance measurements were carried out by contacting the electrode structures with an MPI TS50 manual probe station. Current–voltage characteristics were recorded using a Keithley 2612B source meter, with the applied voltage swept from −5 to 5 V in increments of 0.1 V while monitoring the resulting current. Details of the conductivity measurement were described in detail in SI, section S8.

**Temperature-dependent Transport**: Two-terminal I–V sweeps in the range of ±15 mV were carried out in darkness under high vacuum (base pressure: $10^{-4}$ mbar) using a Keithley 4200A-SCS parameter analyzer. Tungsten probes of 25 μm diameter were positioned on the gold contact pads. For temperature-dependent measurements, the samples were cooled from 25 K to 300 K in a liquid-helium continuous-flow cryostat (base pressure: $10^{-6}$ mbar). Conductance values were extracted by applying a linear fit to the measured I–V data. Section S9 of the SI provides additional context on the analyzed measurements.

**Characterization of Chemiresistive Properties:** Chemiresistive responses of $Ti_3C_2T_x$ MXene-DA composite thin films were screened upon exposure to different VOCs (acetone, toluene, ethanol, 1-propanol, MIBK) and water. Sensor chips containing interdigitated electrodes on soda-lime glass, carrying the composite films, were placed in a custom-built test cell with a chamber volume of ~18 mL. The sensors were alternatingly exposed to nitrogen (5.0) containing the analyte at varying concentrations ranging from 50 to 4000 ppm (analyte gas, AG) for 4 min and to pure nitrogen (5.0) (zero gas, ZG) for 16 min using a custom-built setup. The flow through the test cell was set to 800 mL/min. The sensors' resistances were monitored via applying a constant voltage of 5 V and continuously recording the resulting current using Keithley 2604B and 2602 source measure units. All transients that belong to a flow rate below 10% of our mass flow controllers maximum flow, were considered unreliable and excluded for the evaluation. This refers to 50, 100 and 200 ppm for acetone and 50 ppm for ethanol.

**Computational Methods**

All DFT structure optimizations were performed using the plane-wave technique as implemented in the program package Vienna ab initio simulation package (VASP 6.4.3). An energy cutoff of 500 eV was used for the plane-wave basis set. For the description of the exchange-correlation energy, the functional of Perdew, Burke, and Ernzerhof (PBE)[37] was used in combination with Grimme's empirical dispersion correction (DFT-D3) with a zero-damping function.[38] The structure optimizations were performed using the conjugate gradient algorithm with a convergence threshold for the electronic self-consistent loop of $10^{-5}$ eV. The Brillouin zone was represented by a Monkhorst-Pack k-point mesh of 4 x 4 x 1.

## 5. Acknowledgement

Y.B. and A.F. acknowledge financial support from the Deutsche Forschungsgemeinschaft (DFG) SPP 2100.

H.S., L.M., A.E. and B.R.B. acknowledge financial support funded by the Deutsche Forschungsgemeinschaft (DFG, German Research Foundation) – GRK 2767 – Project number 451785257.

M.A.H. acknowledge the University of Jordan and Leibniz-Institute for Polymer Research Dresden (IPF Dresden) for financial support.

S.S. and Y.V. acknowledge Marie Skłodowska Curie grant agreement No 101066273 for funding this project.

Q.L acknowledge financial support from the Deutsche Forschungsgemeinschaft (DFG).

K.S. and C.H. acknowledge support by the Cluster of Excellence "CUI: Advanced Imaging of Matter" of the Deutsche Forschungsgemeinschaft (DFG) (EXC 2056, funding ID 390715994).

K.S. used the HPC-cluster Hummel-2 at the University of Hamburg for this work. The Cluster was funded by Deutsche Forschungsgemeinschaft (DFG, Funding ID 498394658).

## 6. Conflict of Interest

The authors declare no conflict of interest.

Supporting Information

**Molecular Cross-linking of MXenes: Tunable Interfaces and Chemiresistive Sensing**

*Yudhajit Bhattacharjee,*[1,*] *Lukas Mielke,*[1] *Mahmoud Al-Hussein,*[1,2] *Shivam Singh,*[‡,3,4] *Karen Schaefer,*[‡,5,6] *Borja Rodriguez-Barea ,*[7,8] *Qiong Li,*[1] *Anik Kumar Ghosh,*[1] *Artur Erbe,*[7,8] *Carmen Herrmann,*[5,6] *Yana Vaynzof,*[3,4] *Andreas Fery,*[1,9] *Hendrik Schlicke*[1,*]

[1]Leibniz Institute of Polymer Research Dresden, 01069 Dresden, Germany

[2]Department of Physics, University of Jordan, 11942 Amman, Jordan

[3]Leibniz Institute for Solid State and Materials Research Dresden, 01069 Dresden, Germany

[4]Department of Chemistry, University of Hamburg, 20146 Hamburg, Germany

[5]The Hamburg Centre for Ultrafast Imaging, 22761 Hamburg, Germany

[6]Chair of Emerging Electronic Technologies, Dresden University of Technology, 01187 Dresden, Germany

[7]Institute of Ion Beam Physics and Materials Research, Helmholtz-Zentrum Dresden-Rossendorf (HZDR), 01328 Dresden, Germany

[8]Institute of semiconductors and microsystems, Dresden University of Technology, 01062 Dresden, Germany

[9]Chair of Physical Chemistry of Polymeric Materials, Dresden University of Technology, 01062 Dresden, Germany

‡Equal contribution
*Corresponding authors
Email: bhattacharjee@ipfdd.de, schlicke@ipfdd.de

## S1: Topographic Characterization of Pristine $Ti_3C_2T_x$ via AFM

Topographic AFM measurements (Figure S1a,b) were recorded on isolated flakes from pristine $Ti_3C_2T_x$ material. Therefore, pristine $Ti_3C_2T_x$ was dispersed in ethanol (approx. 0.5 mg/mL) and sonicated for 30 min. The solution was drop-cast and dried overnight on a Si substrate. The thickness of the flakes is approximately 1 nm (cf. the profile scan in Figure S1c).

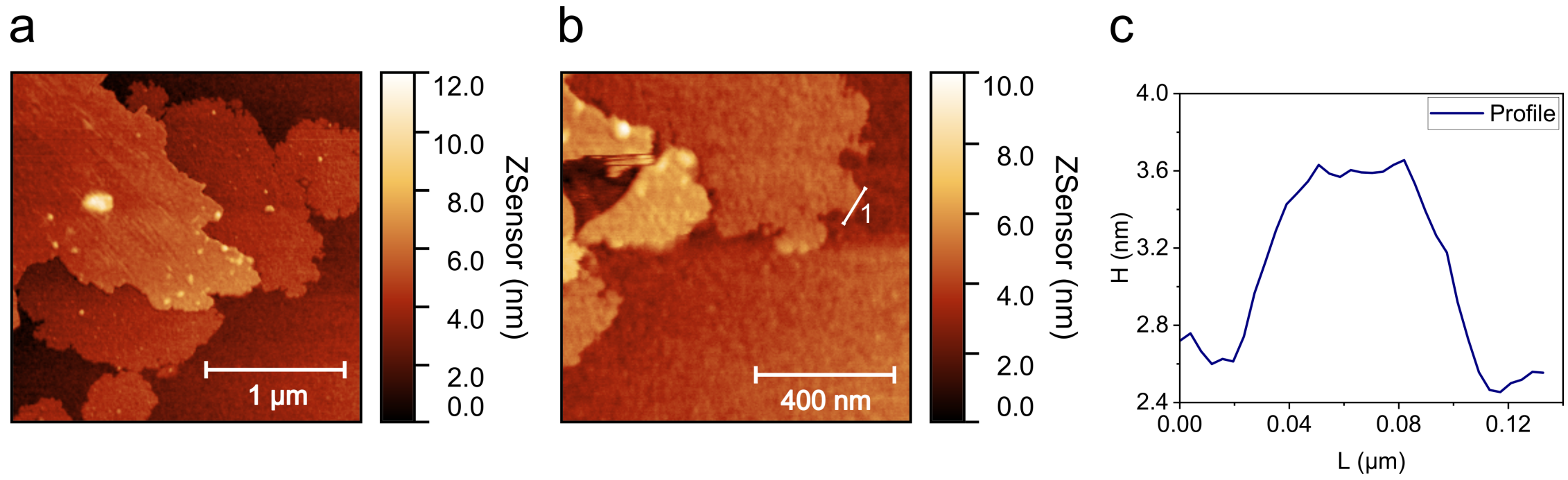

Figure S1. (a,b) Topographic AFM images of neat $Ti_3C_2T_x$ flakes obtained by drop-casting an ethanolic solution onto a wafer substrate. (c) Height profile extracted from AFM scan shown in b, as indicated.

## S2: Synthesis of OAm-Functionalized MXene Dispersion

First, commercially supplied pristine $Ti_3C_2T_x$ MXene powder (Figure 1b and S1) from Nanochemzone, Inc. Canada, was dispersed in chloroform (10 mg/mL), and oleylamine (55 mM) was added. The resulting dispersion was sonicated for one hour and kept stirring for around 48 h at room temperature. After that, the dispersion was again sonicated for about an hour before purification. The resulting $Ti_3C_2T_x$–OAm dispersion was subjected to slow centrifugation at 1000 rcf for 1 min to remove any larger agglomerates. The resulting supernatant was isolated and subjected to fast centrifugation at 8000 rcf for 8 min to precipitate the OAm-functionalized MXene sheets. Excess OAm ligands were then removed by discarding the supernatant. Subsequently, the precipitate was redispersed in chloroform. This resulting dispersion underwent another slow centrifugation step at 1000 rcf for 1 min to eliminate any residual agglomerate. The resulting supernatant was carefully transferred into a fresh tube and sonicated for 30 min to achieve a homogeneously dispersed solution. This final dispersion was used as the spin-coating stock solution for thin film preparation.

### S3: Stability of the Colloidal Dispersion of OAm Functionalized $Ti_3C_2T_x$

Figure S2 depicts a stable dispersion of OAm functionalized $Ti_3C_2T_x$ obtained by the procedure described in S2. The dispersion, which was stored in a refrigerator (4-6 °C) is stable over a period of 30 days and beyond, without any visible precipitation.

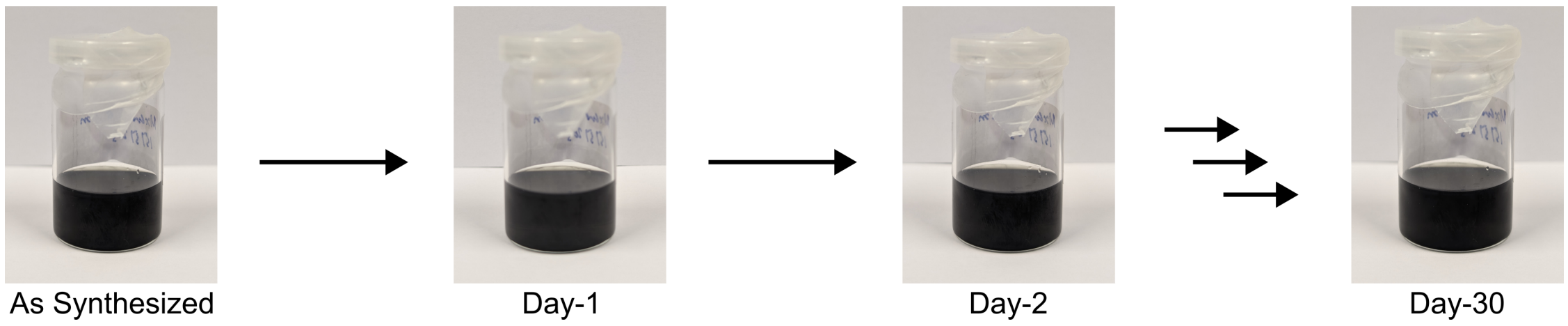

Figure S2. Photographs depicting a stable dispersion of OAm Functionalized $Ti_3C_2T_x$ over a period of 30 day.

### S4: Charge Transport in Composite Films from OAm functionalized $Ti_3C_2T_x$

Current–voltage (I–V) characteristics of a thin film formed by depositing an $Ti_3C_2T_x$-OAm dispersion via spin coating at 1000 rpm onto an interdigitated electrode (IDE) with a channel length (distance between electrode fingers) of 20 µm and an overlap width of 49 mm.

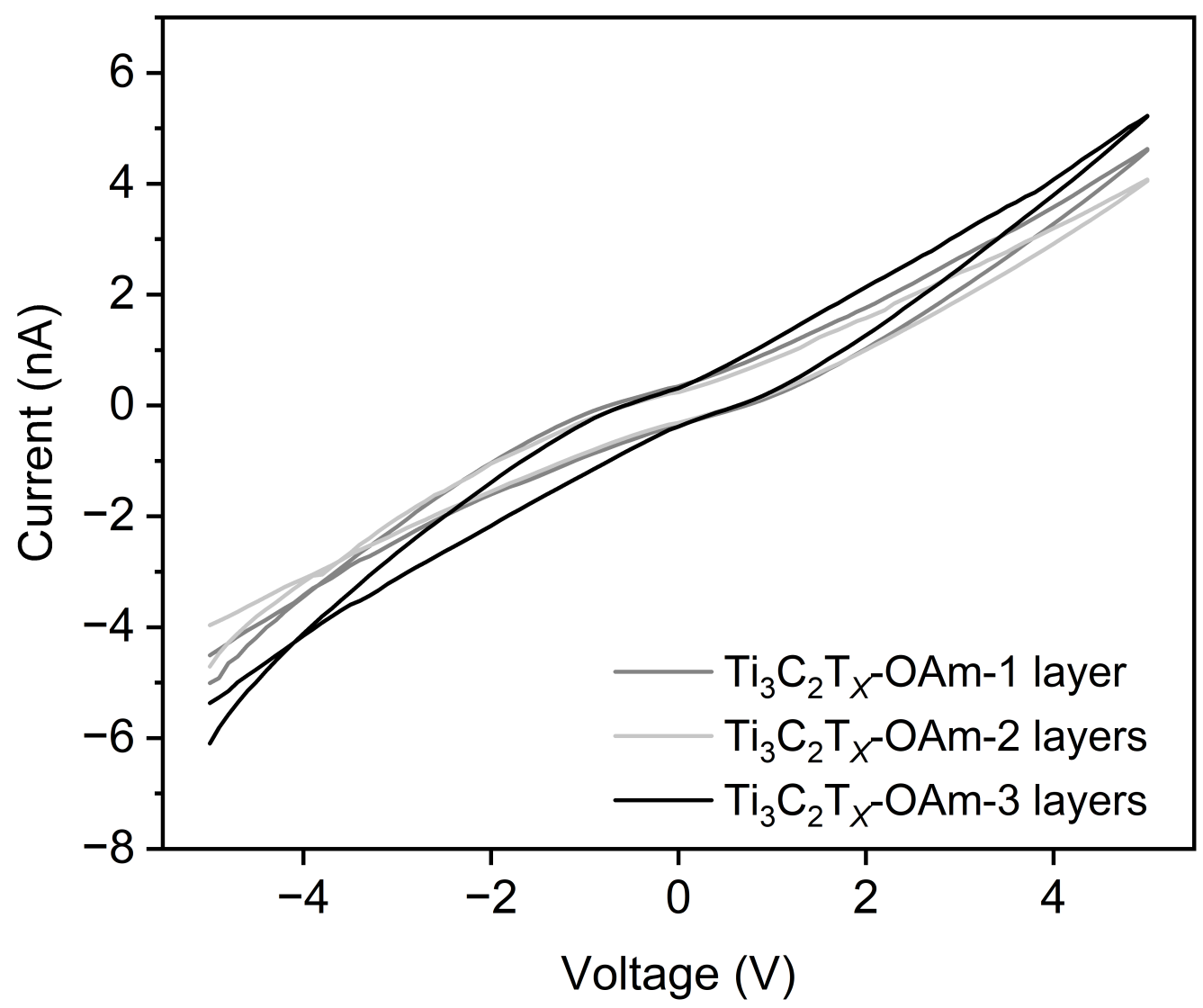

Figure S3. Current–voltage (I–V) characteristics of the film composed of $Ti_3C_2T_x$-OAm dispersion fabricated via spin coating.

### S5: Reproducibility of Device Fabrication Using Aged $Ti_3C_2T_x$–OAm Dispersions

$Ti_3C_2T_x$–6DA / 8DA / 10DA devices having a film thickness of ~(32, 30 and 31) nm respectively were prepared in similar way as described in section 2.6 in the main manuscript, using interdigitated electrode structures (IDEs) with a channel length (distance between electrode fingers) of 20 µm and an overlap width of 49 mm. The devices made from the 30 day-old MXene–OAm dispersion, did not show any pronounce change in current output.

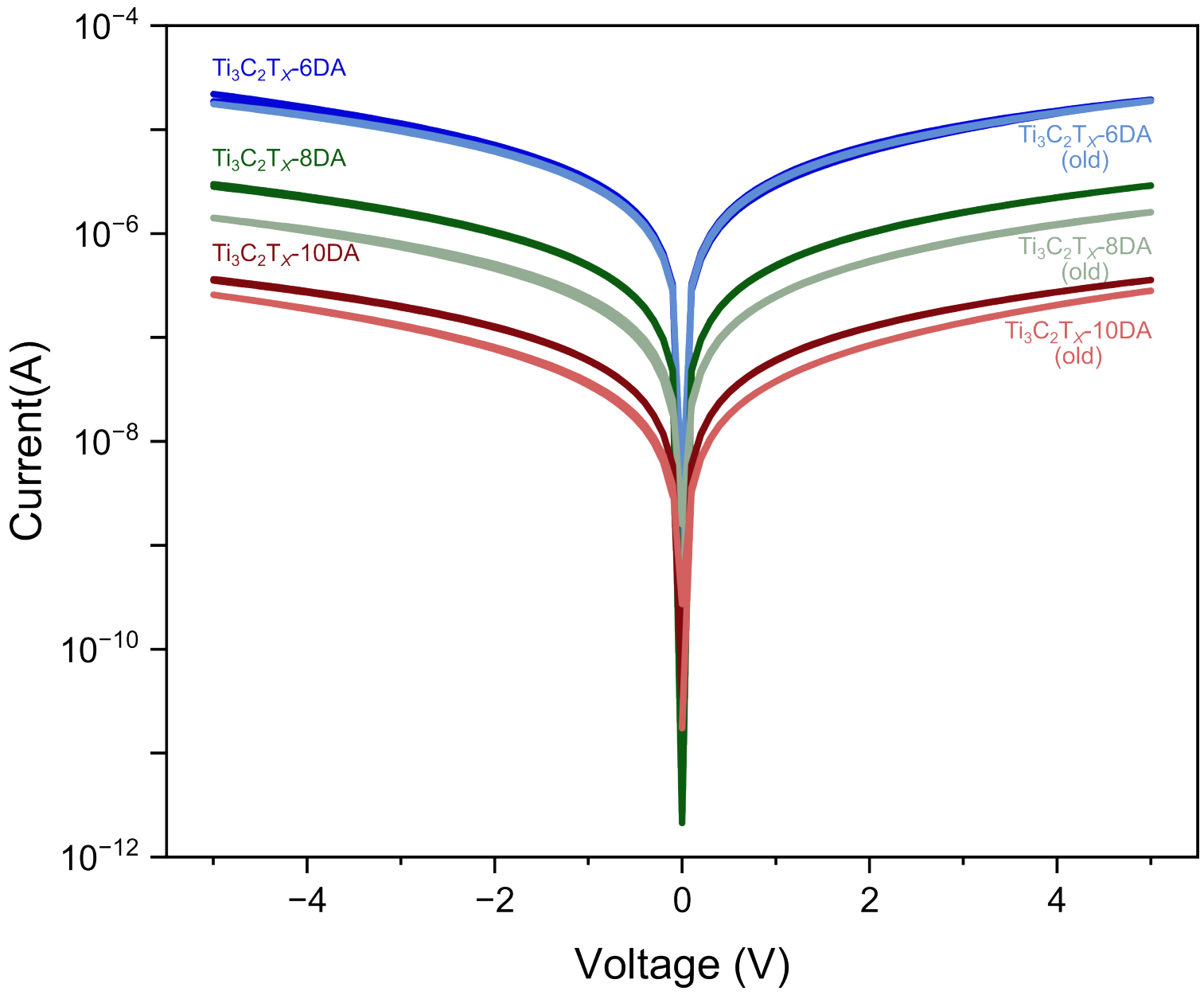

Figure S4. Current–voltage (I–V) characteristics of the $Ti_3C_2T_x$-DA devices fabricated using freshly prepared and aged $Ti_3C_2T_x$–OAm dispersions.

### S6: First Principles Calculations

Table S1. Adsorption energies of a butyl amine ligand on $Ti_3C_2(OH)_2$, $Ti_3C_2F_2$, $Ti_3C_2T_x$ monolayers.

| Species | $E_{ads}$ (kJ/mol) |
|---|---|
| $Ti_3C_2(OH)_2$ | -33.8 |
| $Ti_3C_2F_2$ | -21.2 |
| $Ti_3C_2T_x(Ti_3C_2F(OH))$ | -51.4 |

## S6.1: Details on the Preparation of the Monolayer Structures for the Investigation of the Adsorption of Amine Ligands on MXene Monolayers

Unit cell structures of $Ti_3C_2(OH)_2$ and $Ti_3C_2F_2$ were taken from DFT structure optimizations by Tang et al.[1] using a Perdew, Burke, and Ernzerhof (PBE) functional. From these unit cells, $3 \times 3$ supercells were constructed that were used as monolayers for the investigation of the adsorption of amine ligands on $Ti_3C_2T_x$ surfaces (see Section 2.4.1 of the main manuscript). This unit cell size ensured that the monolayer surface was sufficiently large to prevent artificial interactions between amine ligands under periodic boundary conditions. To create the $Ti_3C_2F(OH)$ monolayer, we started with the $Ti_3C_2(OH)_2$ structure and replaced half of the -OH groups with -F groups. Then, the resulting super cell were optimized using DFT with a PBE functional. For computational details see the computational methods section of the main manuscript. The optimized structures of the diamine ligand on the $Ti_3C_2(OH)_2$ and $Ti_3C_2F_2$ monolayers are shown in Figure S5.

## S6.2: Details on the preparation of the structures and the protocol for finding the optimal center-to-center distance of Mxene monolayers

The MXene monolayers were formed by creating a $2 \times 2$ supercell from the unit cells of each layer type $Ti_3C_2(OH)_2$, $Ti_3C_2F_2$. The $Ti_3C_2F(OH)$ structure was additionally created by replacing half of the -OH groups of the $Ti_3C_2(OH)_2$ monolayer with -F groups. The monolayers of each MXene type were optimized using DFT and a PBE functional. Then, a diamine ligand was adsorbed onto each of the three optimized MXene structures. The three monolayer configurations were optimized by fixing the positions of the atoms in the MXene monolayer, allowing for the optimization of the bond distances between the diamine and the MXene surface. The diamine ligands were initially straight and maintained this non-tilted conformation after the structural optimization. Finally, the second MXene monolayer, optimized as before, was placed on top of the free amine group of the diamines, with the same N-to-layer distance as the bottom MXene monolayer. The resulting two-layered structure was then optimized, again keeping the atoms in the MXene monolayers frozen.

To determine the optimal center-to-center distance between the two monolayers, the initially optimized two-layer structure was modified by decreasing the separation between the layers by 0.2 Å through a shift of the atomic coordinates. The modified structure was then re-optimized while keeping the atoms within the MXene layers fixed. The total energy of this configuration was compared to that of the structure prior to the shift. If the energy was lower, the distance between the layers was further reduced by 0.2 Å. This iterative process was repeated until two consecutive reductions no longer resulted in a decrease in total energy, indicating that the optimal separation had been reached.

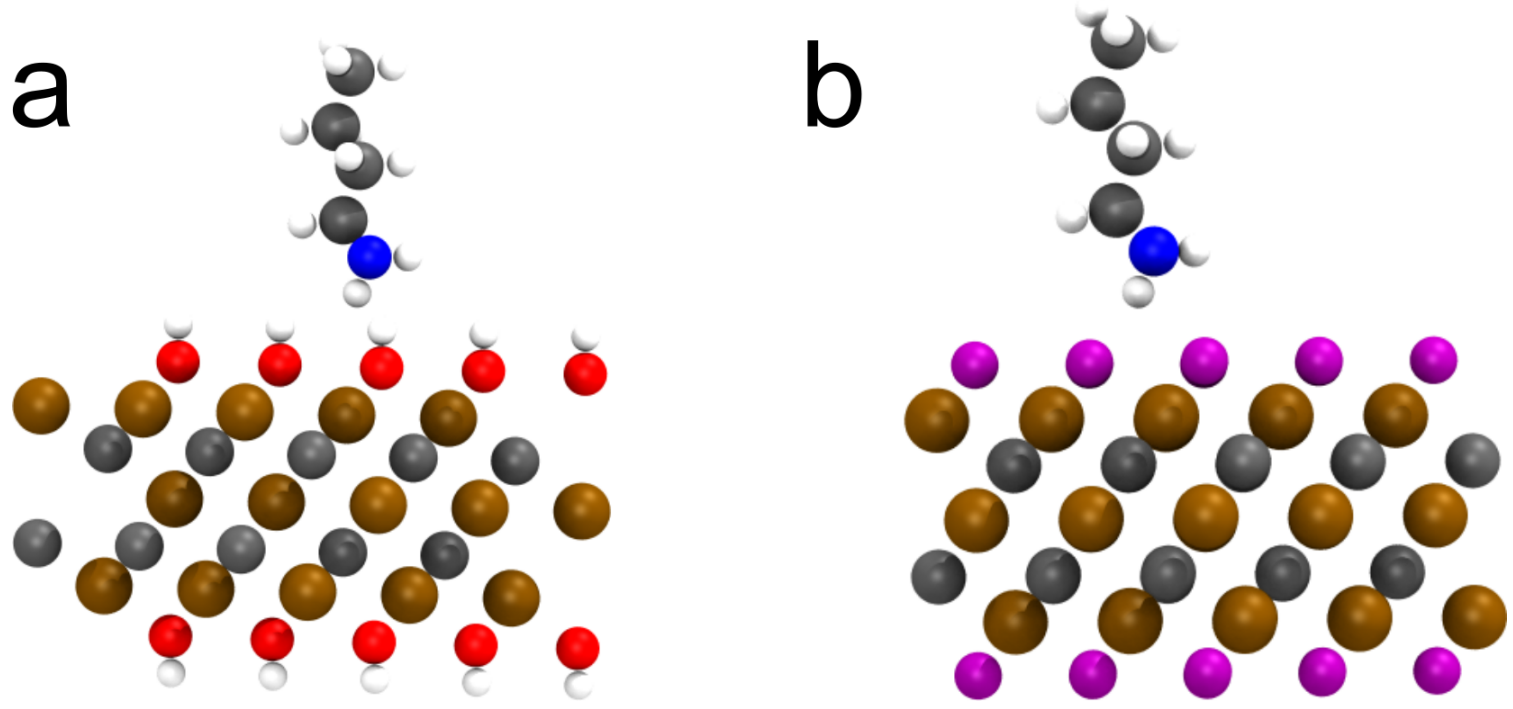

Figure S5. Optimized structures of butyl amine ligands adsorbed on (a) $Ti_3C_2(OH)_2$ (b) and $Ti_3C_2F_2$.

### S7: XPS of Ti 2P and O 1s

Figure S6 shows the XPS spectra of Ti 2P and O 1S of various $Ti_3C_2T_x$ -DA composites.

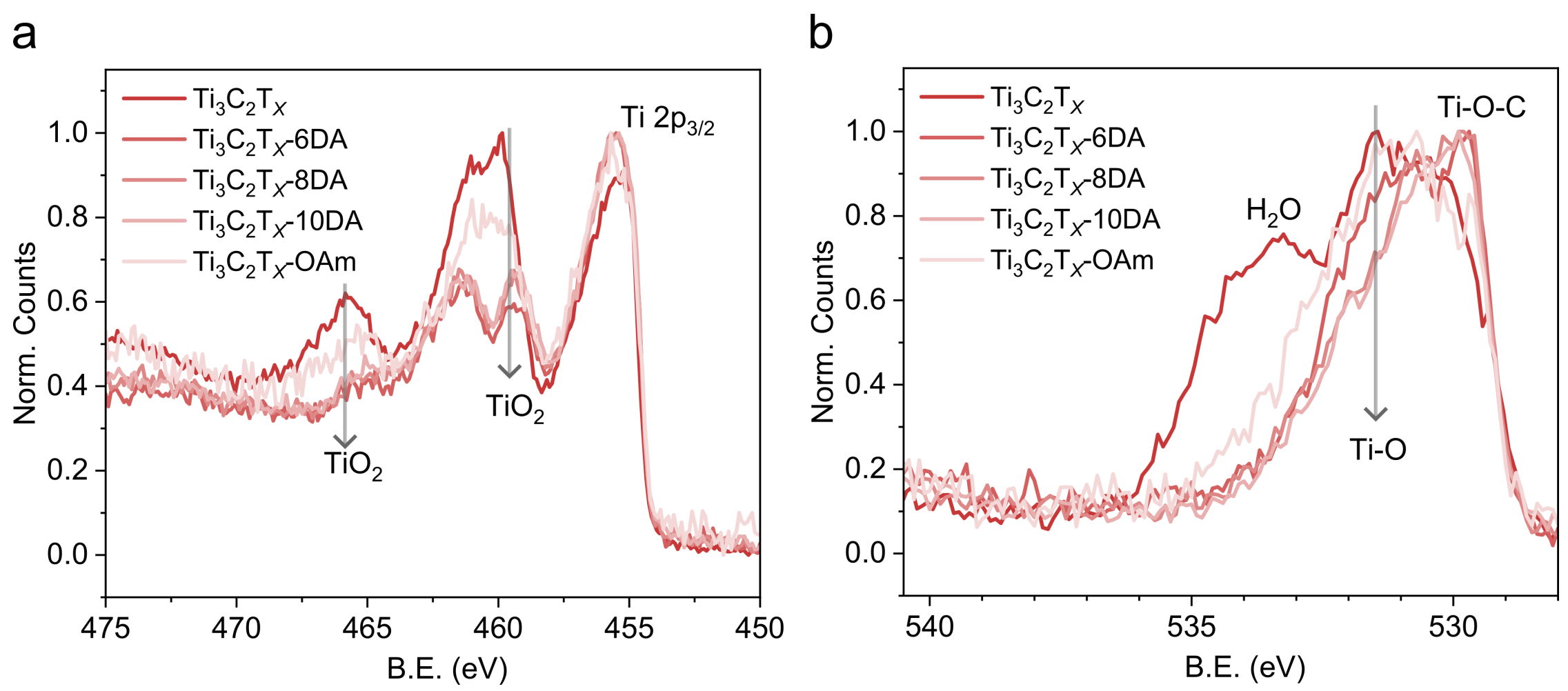

Figure S6. The XPS core level spectra of (a) Ti 2p and (b) O 1s for$Ti_3C_2T_x$, $Ti_3C_2T_x$–6DA, $Ti_3C_2T_x$–8DA, $Ti_3C_2T_x$–10DA, and $Ti_3C_2T_x$–OAm.

### S8: Film Thickness and Conductivity Determination

AFM scans were carried out at the edges of scratches deliberately introduced into $Ti_3C_2T_x$-DA composite films deposited onto wafers equipped with electrode structures. The recorded data were first corrected for tilt and artefacts using the software Gwyddion 2.63. Subsequently, film, substrate, and edge (excluded) areas were manually selected, as exemplarily shown in Figure S7. A custom Python script was utilized to determine the film thickness t as the mean height value of the selected film area. Figure S7b–d shows exemplary scans of 6DA (b), 8DA (c), and 10DA (d) cross-linked $Ti_3C_2T_x$ films. Three different positions per sample were imaged. The mean thicknesses obtained were 28 nm, 27 nm, and 26 nm, respectively. It is noteworthy that the films exhibit significant surface roughness. The conductivity σ was calculated according to $\sigma = \frac{G \cdot l}{w \cdot t}$, where $G$ denotes the conductance derived from slope fits to $I$–$V$ curves in the range of −5 V to +5 V, $l$ is the channel length (distance between electrode fingers, $l = 20$ µm), $w$ is the effective channel width (total overlap width of the interdigitated electrode fingers, $w = 49$ mm), and $t$ is the mean film thickness derived from topographic AFM scans. The conductivities of the 6DA (b), 8DA (c), and 10DA (d) cross-linked $Ti_3C_2T_x$ films were $6 \times 10^{-4}$ S/cm, $8 \times 10^{-5}$ S/cm, and $1 \times 10^{-5}$ S/cm, respectively.

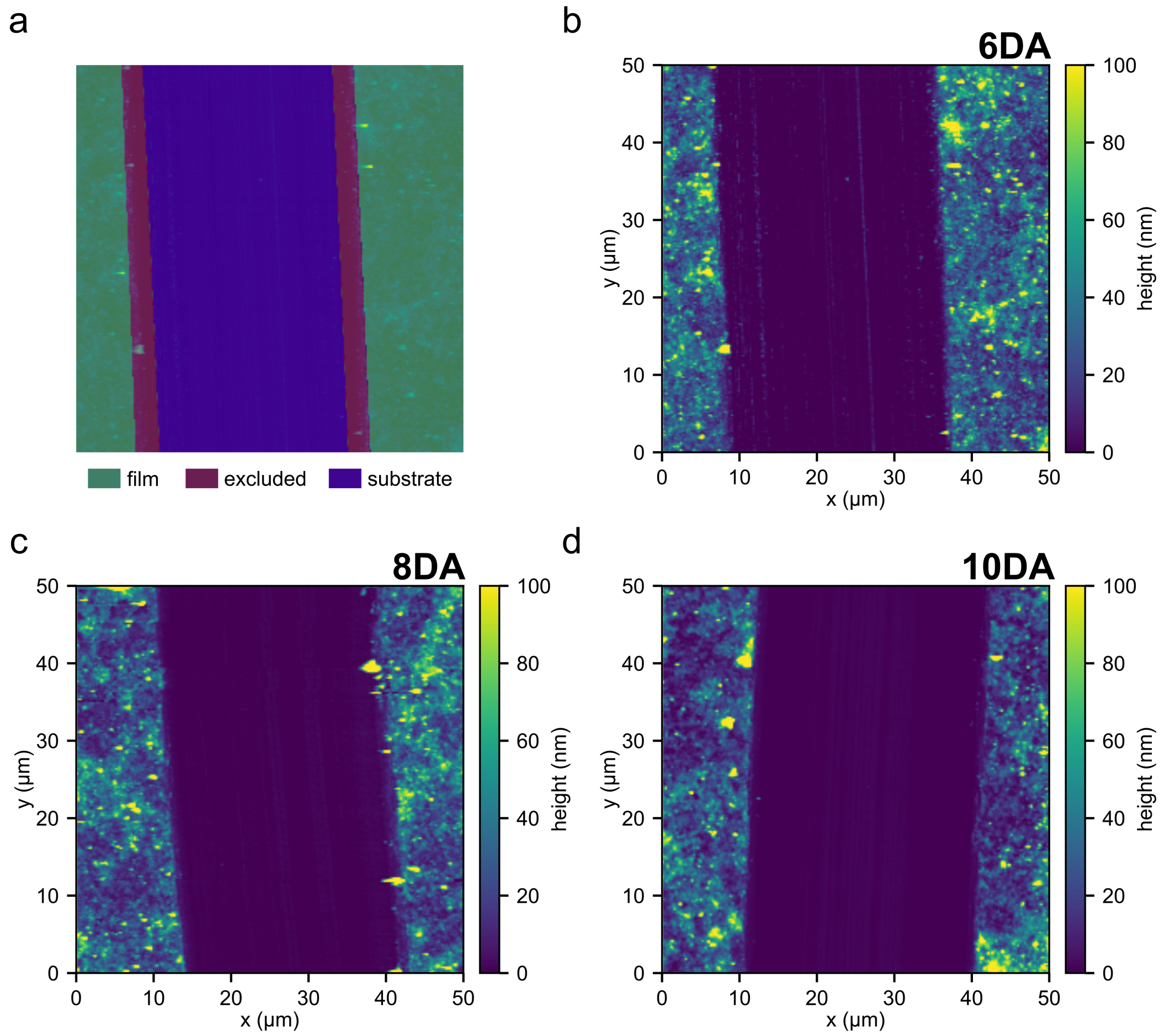

Figure S7. (a) Exemplary mask pattern applied for selecting film and substrate areas for calculations of the film thickness. (b-d) Representative topographic AFM scans of films composed of $Ti_3C_2T_x$ flakes cross-linked with 6DA, 8DA, and 10DA, fabricated via the layer-by-layer spin coating procedure.

## S9: Temperature-Dependent Transport in Functionalized $Ti_3C_2T_x$ Films

For the charge transport study presented in the main text (section 2.6), we measured two sets of temperature-dependent current-voltage (I-V) characteristics on different IDE regions of the same 6DA-, 8DA-, and 10DA-functionalized MXene films. The applied bias was limited to ±15 mV to avoid Joule heating effects, and the resulting IV curves remained highly linear across the entire sweep range (Figure S8). Conductance values were extracted from the slope of the linear fits and used to evaluate the charge transport mechanism of the films (Figure S9). Supplementary measurements confirm the reproducibility of the temperature-dependent electronic properties across different IDE regions.

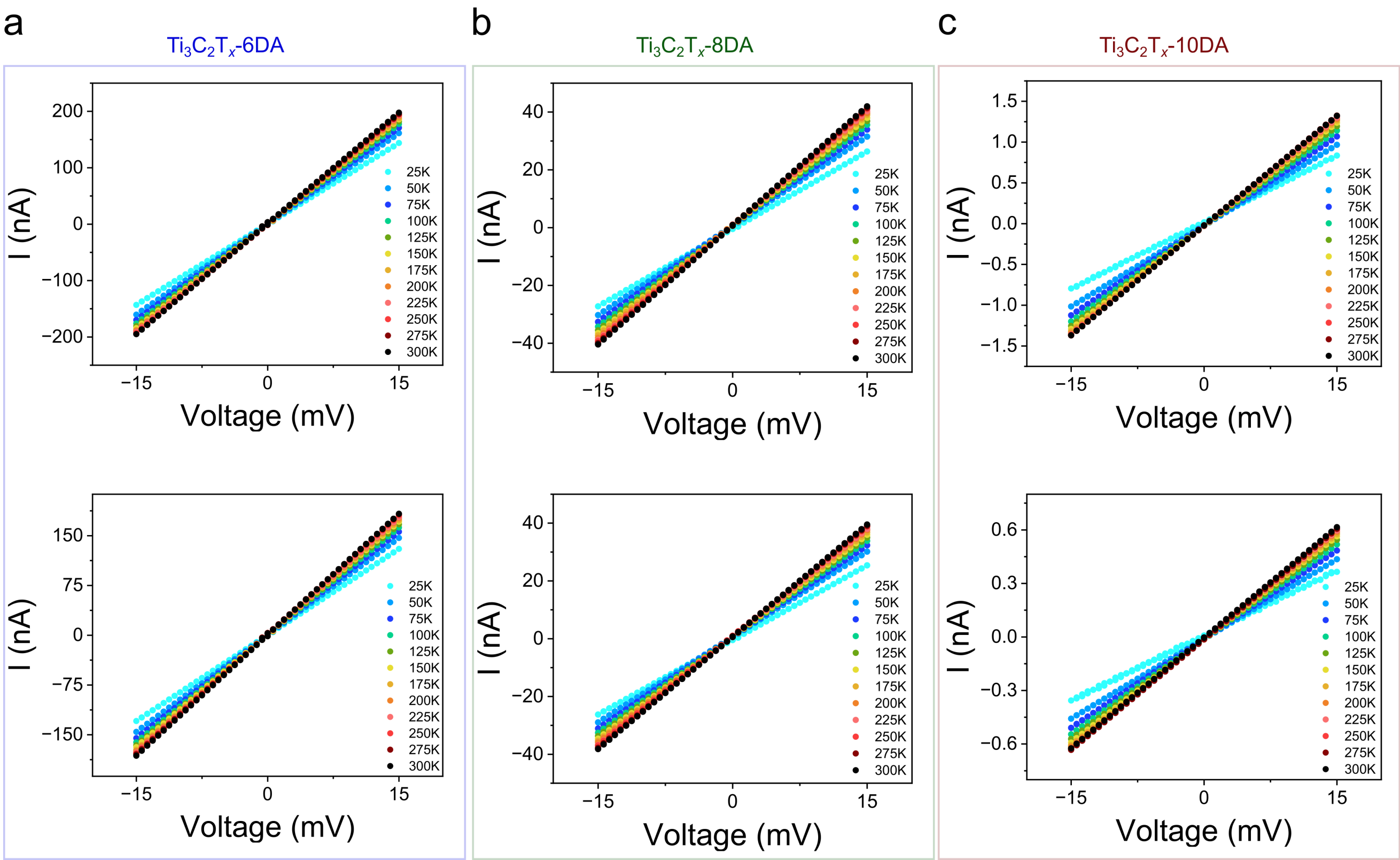

Figure S8. Temperature dependent current–voltage (I–V) characteristics of $Ti_3C_2T_x$ films functionalized with 6DA, 8DA, and 10DA ligands, measured across the 25-300 K range under high vacuum and in darkness for ±15 mV voltage sweeps.(a-c) correspond to the IDE devices analyzed in the main text, while (d-f) show the additional devices included in the Supplementary Information.

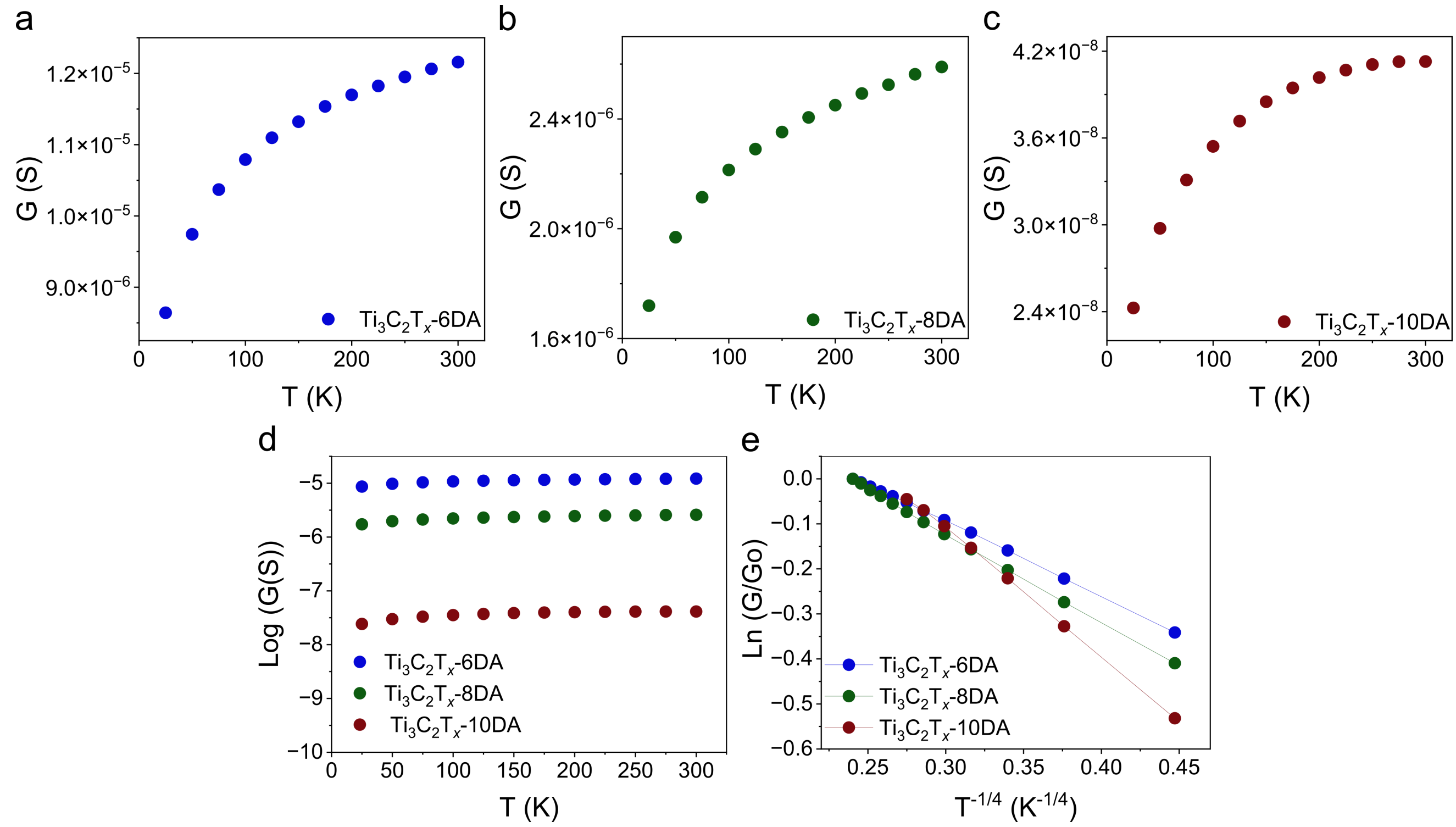

Figure S9. Second set of measured thin films. (a) Top row corresponds to the conductance as a function of temperature for each ligand length. (b) Logarithmic representation of the conductance for the various functionalized $Ti_3C_2T_x$ films. (c) Normalized natural logarithm of the conductance versus $\frac{1}{T^{1/4}}$ for each functionalized film. Solid lines indicate linear fits to the experimental data; for the 10DA-functionalized film the fit is restricted to the 25-200 K range.

## S10: Stability of the $Ti_3C_2T_x$-6DA Device in Ambient Atmosphere

The $Ti_3C_2T_x$–6DA chemiresistor device (having specifications described in S5) was used for multiple VOC sensing experiments and stored under ambient conditions. Its I–V characteristics were periodically measured under a nitrogen flow. For more than 17 days, the device did not exhibit any significant change in its I–V response. After 6 months, a noticeable drop in current was observed, although the device still maintained appreciable conductivity.

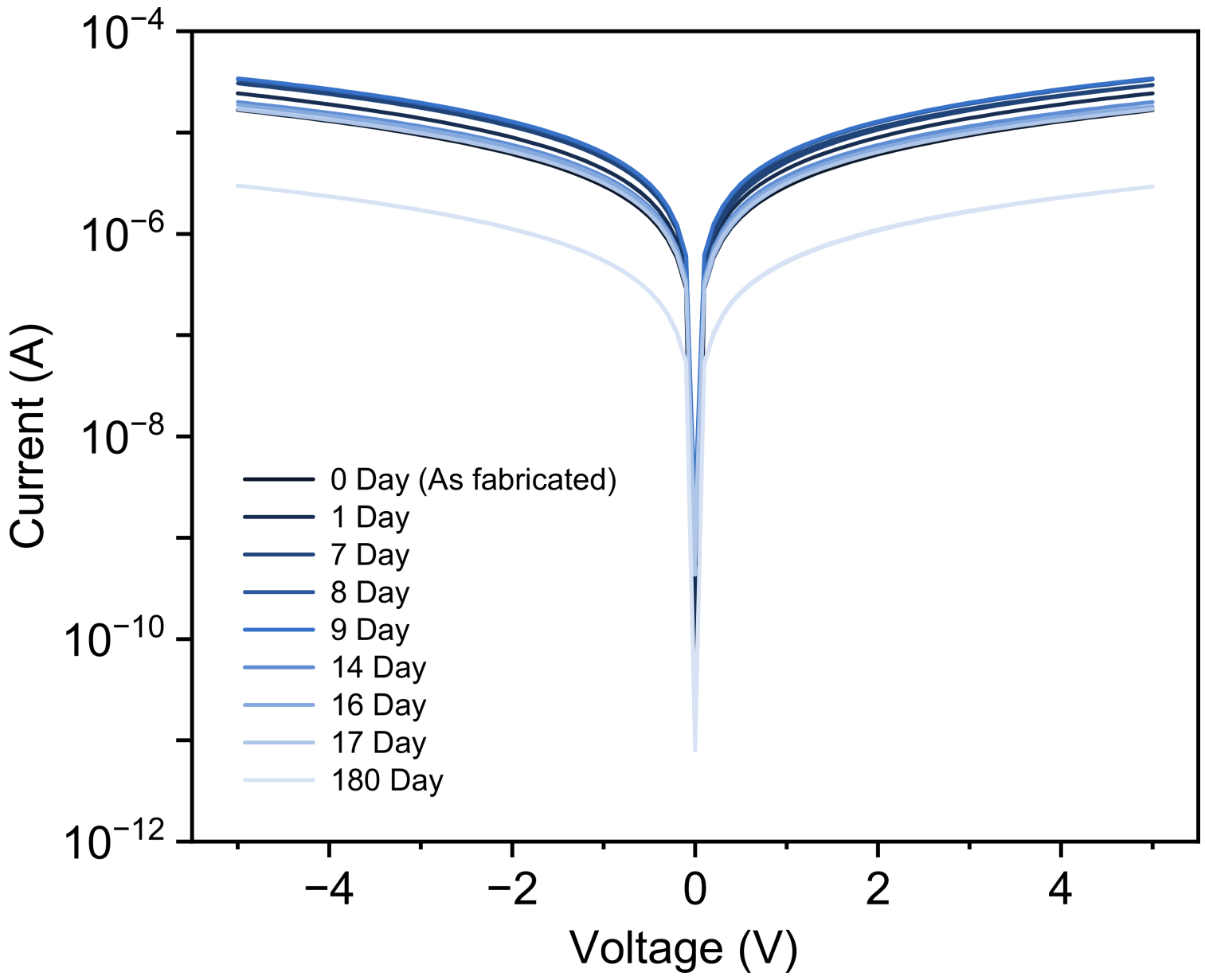

Figure S10. Current–voltage (I–V) characteristics of the $Ti_3C_2T_x$-6DA device kept under ambient atmosphere for 6 months and subjected to multiple chemiresistive VOC sensing experiment.

## S11: Resistive VOC Sensing

The following figures show all baseline corrected responses towards different gaseous analytes of the neat $Ti_3C_2T_x$ (Figure S11), $Ti_3C_2T_x$-6DA (Figure S12) and $Ti_3C_2T_x$-10DA composites (Figure S13). Figure S14a and b show exemplary resistive responses of a 10DA cross-linked $Ti_3C_2T_x$ thin film exposed to toluene and water vapor with concentrations ranging from 50 – 4000 ppm. Upon exposure, the sensors showed a fully reversible increase of resistance of up to 26% (water vapor, 4000 ppm) with a short response time. Notably, in comparison to the $Ti_3C_2T_x$-6DA device shown in the main article, the $Ti_3C_2T_x$-10DA device did not show a visible response to toluene. This could be due to the overall lower conductance of the device and a resulting lower signal to noise ratio. Figure S14 depicts response isotherms of the $Ti_3C_2T_x$-10DA sensor exposed to acetone, toluene, ethanol, 1-propanol, MIBK and water vapor at varying concentrations. The sensor's resistive sensitivity to the different VOCs was extracted from slope fits to data points in the low-concentration range (50-400 ppm). For acetone the data points corresponding to 50, 100 and 200 ppm, and for ethanol 50 ppm were excluded. Further, based on the isotherms' shape we tentatively fitted a Langmuir-Henry model. The extracted fit parameters are provided in the section S12.

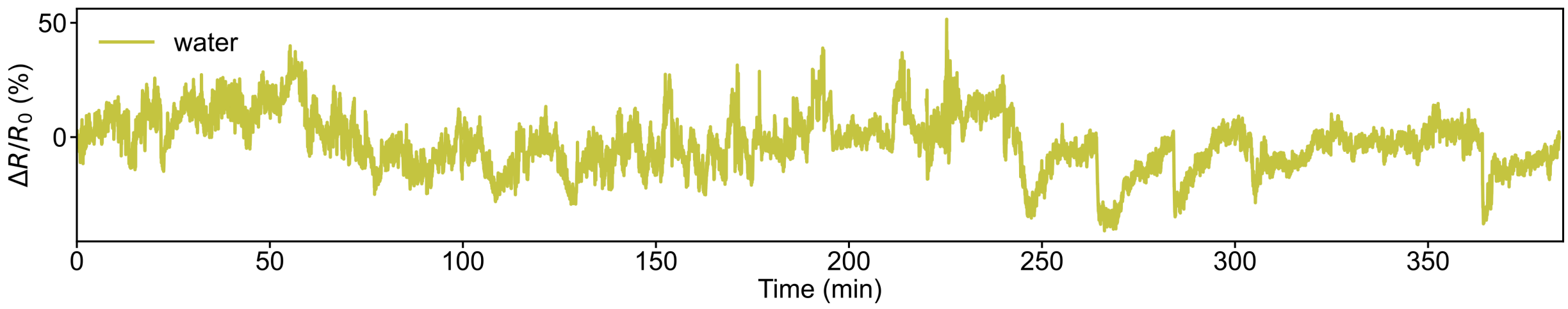

Figure S11. Baseline corrected responses of a neat $Ti_3C_2T_x$ chemiresistor to water. The chemiresistor was transiently exposed to concentrations of 0, 50, 100, 200, 400, 800, 800, 400, 200, 100, 50, 500, 1000, 2000, 4000, 4000, 2000, 1000 and 500 ppm.

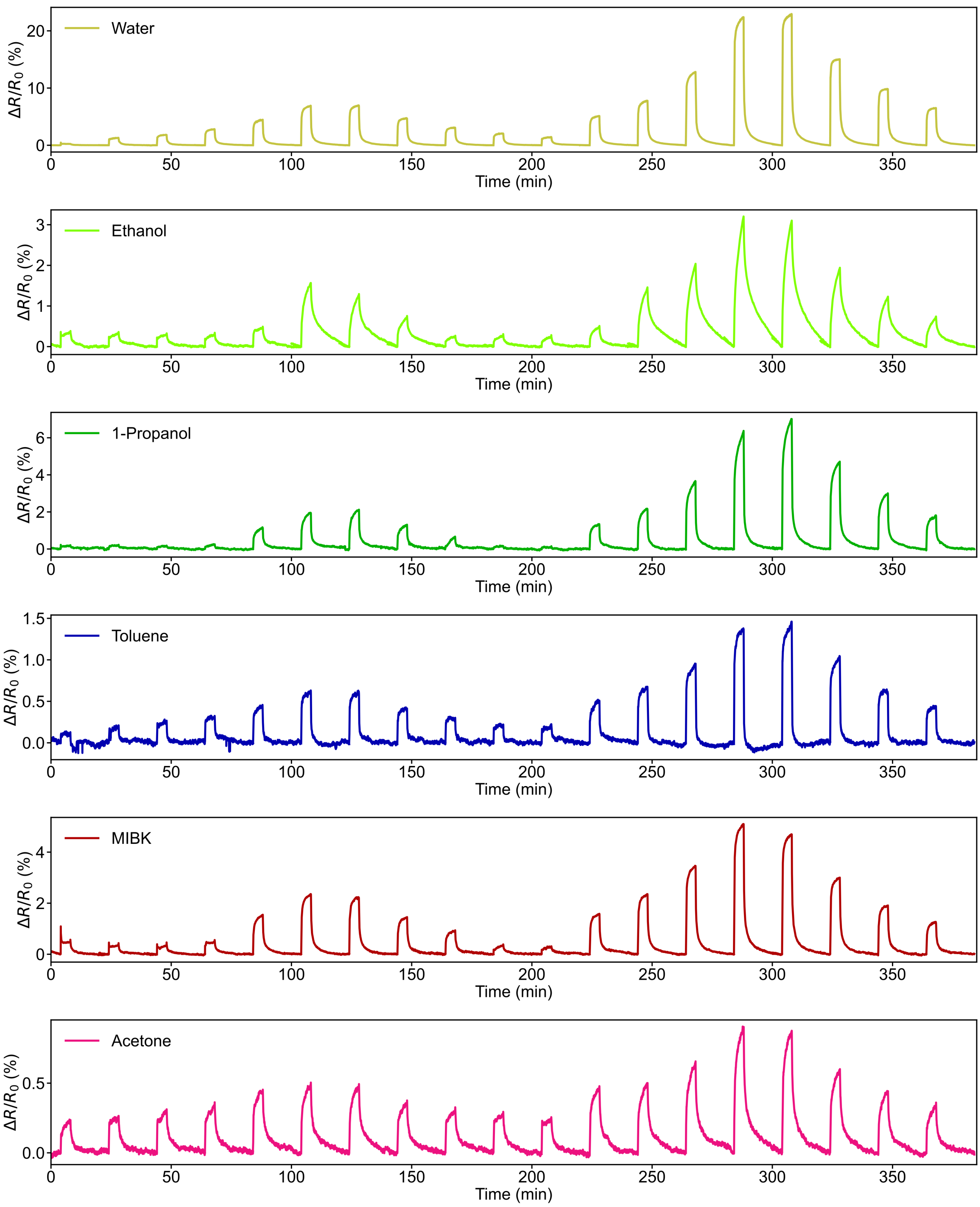

Figure S12. Baseline corrected responses of a $Ti_3C_2T_x$-6DA chemiresistor to water, ethanol, 1-propanol, toluene, MIBK and acetone. The transients correspond to concentrations 0, 50, 100, 200, 400, 800, 800, 400, 200, 100, 50, 500, 1000, 2000, 4000, 4000, 2000, 1000 and 500 ppm.

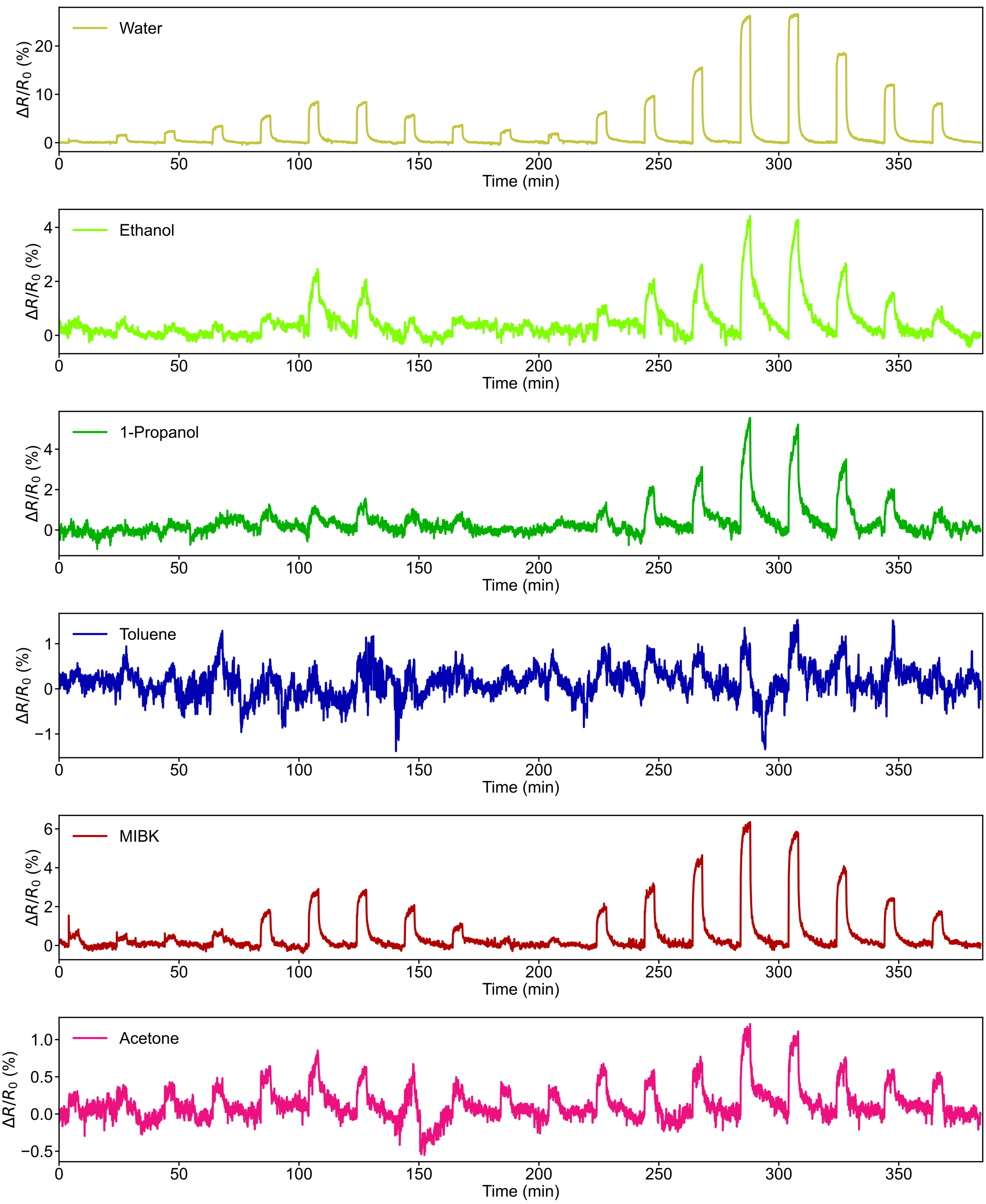

Figure S13. Baseline-corrected responses of a $Ti_3C_2T_x$–10 DA chemiresistor to water, ethanol, 1-propanol, toluene, MIBK, and acetone. The transient signals correspond to concentrations of 0, 50, 100, 200, 400, 800, 800, 400, 200, 100, 50, 500, 1000, 2000, 4000, 4000, 2000, 1000, and 500 ppm.

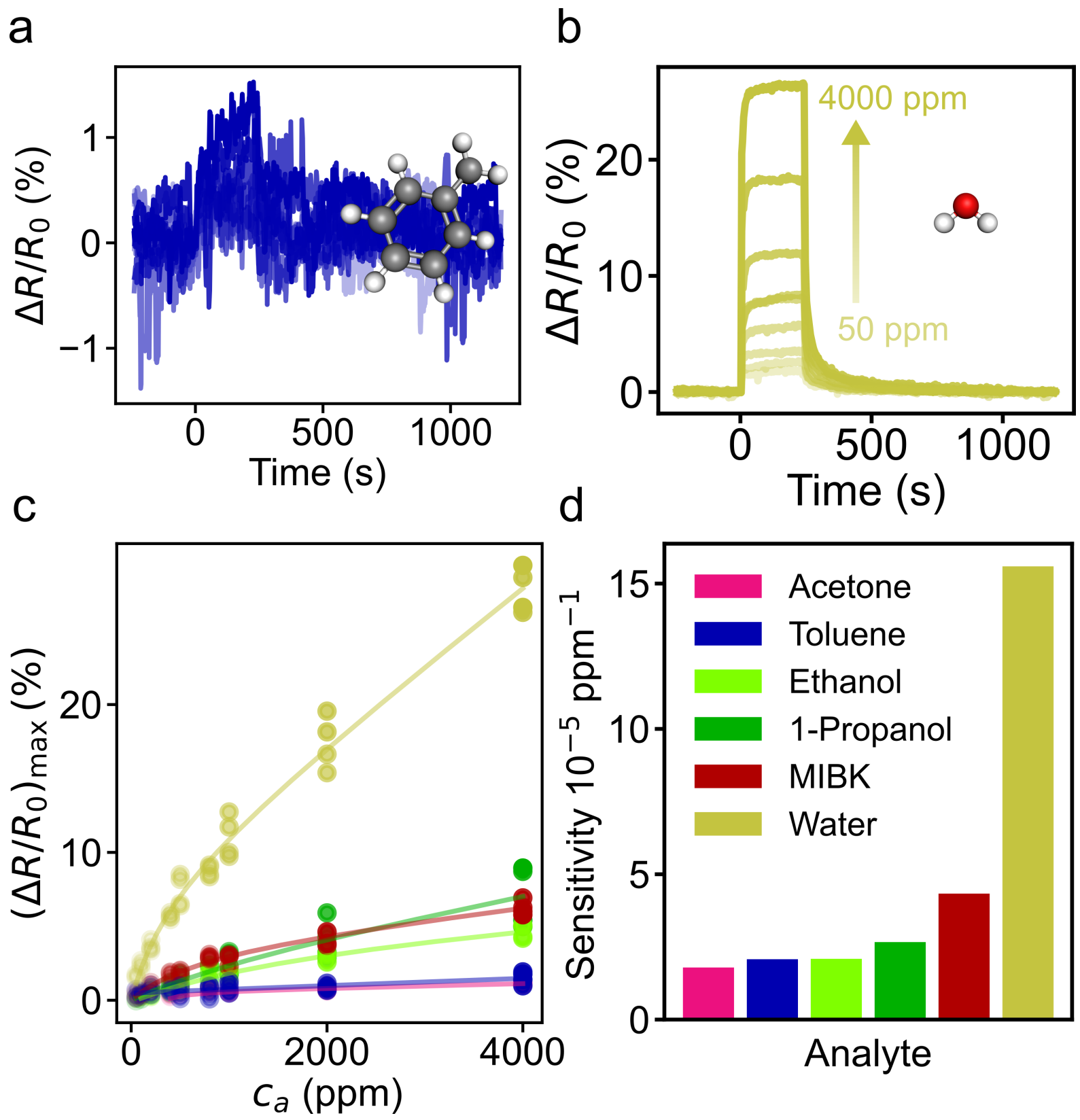

Figure S14. (a, b) Transient response of a $Ti_3C_2T_x$-10DA chemiresistor to (a) toluene and (b) water at concentrations of 50, 100, 200, 400, 500, 800, 1000, 2000 and 4000 ppm in nitrogen. (c) Response isotherms of a $Ti_3C_2T_x$-10DA based chemiresistor. The colors correspond to the probed analyte molecules as shown in d. Through lines depict Langmuir-Henry isotherms fitted to the data. (d) Chemiresistive sensitivities of a $Ti_3C_2T_x$-10DA sensor to test gases, extracted from slope fits in a concentration range of 50-400 ppm. For acetone, just the data point of 400 ppm and for ethanol 400, 200 and 100 ppm were considered to be reliable.

## S12: Langmuir–Henry Model Fit Parameters

The data in Figure 7d and Figure S14c of the main document were fitted using the Langmuir–Henry model, following Equation S1, where $K_b$ is the Langmuir binding constant and $K_h$ is the Henry sorption constant. Sorption constants were calculated assuming standard conditions (T = 298.15 K, p = 1.013 $\times 10^5$ Pa). Notably, the Langmuir-Henry model did not yield an acceptable fit for acetone .

$$\frac{\Delta R}{R_0} = \left(\frac{\Delta R}{R_0}\right)^{\text{sat}} \cdot \frac{K_b \cdot c_A}{1 + K_b \cdot c_A} + K_h \cdot c_A \tag{S1}$$

Table S2. Extracted parameters for different analytes based on the Langmuir-Henry adsorption models for $Ti_3C_2T_x$-6DA system.

| Analyte | $(\Delta R/R_0)^{\text{sat}}$ (%) | $K_b$ (L/mol) | $K_h$ (L/mol) |
|---|---|---|---|
| 1-Propanol | 5.90 | $10.5 \times 10^3$ | 165 |
| Acetone | -112 | $12.9 \times 10^{-3}$ | 65.0 |
| Ethanol | 5.24 | $75.2 \times 10^2$ | 22.4 |
| MIBK | 2.43 | $44.9 \times 10^3$ | 164 |
| Toluene | 0.49 | $17.4 \times 10^4$ | 67.8 |
| Water | 5.27 | $79.9 \times 10^3$ | $10.1 \times 10^2$ |

Table S3. Extracted parameters for different analytes based on the Langmuir-Henry adsorption models for $Ti_3C_2T_x$-10DA system.

| Analyte | $(\Delta R/R_0)^{\text{sat}}$ (%) | $K_b$ (L/mol) | $K_h$ (L/mol) |
|---|---|---|---|
| 1-Propanol | 3.03 | $14.5 \times 10^3$ | 299 |
| Acetone | 0.63 | $10.2 \times 10^4$ | 32.1 |
| Ethanol | 4.35 | $12.2 \times 10^3$ | 105 |
| MIBK | 3.81 | $33.5 \times 10^3$ | 183 |
| Toluene | 0.50 | $40.6 \times 10^5$ | 59.9 |
| Water | 8.03 | $60.9 \times 10^3$ | $12.6 \times 10^2$ |